\documentclass[1p]{elsarticle}         

\usepackage{amssymb}
\usepackage{amsmath}
\usepackage[linesnumbered,ruled,algo2e]{algorithm2e}
\usepackage{graphicx}
\usepackage{hyperref}
\usepackage{booktabs}
\usepackage{multicol}
\usepackage{multirow}
\usepackage{rotating}
\usepackage{mathrsfs}
\usepackage{microtype}
\usepackage{subcaption}
\usepackage{tikz}
\usepackage{pgfplots}

\usepackage{placeins} 

\usepackage{array}
\newcolumntype{P}[1]{>{\raggedleft\arraybackslash}p{#1}}

\long\def\ignore#1{}

\let\myclearpage=\clearpage
\def\myclearpage{\FloatBarrier\clearpage}

\let\myclearpage=\relax     

\let\AJPii=\ignore

\usepackage{color}

\makeatletter
\DeclareRobustCommand{\varname}[1]{\begingroup\newmcodes@\mathit{#1}\endgroup}
\makeatother

\makeatletter
\g@addto@macro{\UrlBreaks}{\UrlOrds}
\makeatother

\title{Combining Monte-Carlo and Hyper-heuristic methods for the Multi-mode Resource-constrained Multi-project Scheduling Problem}

\author[nott]{Shahriar~Asta}
\ead{sba@cs.nott.ac.uk}

\author[nott,essex]{Daniel~Karapetyan\corref{cor1}}
\ead{daniel.karapetyan@gmail.com}

\author[nott]{Ahmed~Kheiri}
\ead{axk@cs.nott.ac.uk}

\author[nott]{Ender~\"{O}zcan}
\ead{exo@cs.nott.ac.uk}

\author[nott]{Andrew~J.~Parkes}
\ead{ajp@cs.nott.ac.uk}

\address[nott]{University of Nottingham, School of Computer Science\\ 
Jubilee Campus, Wollaton Road, Nottingham, NG8 1BB, UK}

\address[essex]{University of Essex, Institute for Analytics and Data Science\\ 
Wivenhoe Park, Colchester, CO4 3SQ, UK}

\cortext[cor1]{Corresponding author}

\date{\today}

\begin{document}

\begin{abstract}
\begin{sloppypar}
 Multi-mode resource and precedence-constrained project scheduling is a well-known challenging real-world optimisation problem. 
 An important variant of the problem requires scheduling of activities for multiple projects considering availability of local and global resources while respecting a range of constraints. 
 A critical aspect of the benchmarks addressed in this paper is that the primary objective is to minimise the sum of the project completion times, with the usual makespan minimisation as a secondary objective. 
 We observe that this leads to an expected different overall structure of good solutions and discuss the effects this has on the algorithm design.
 This paper presents 
 a carefully designed hybrid of Monte-Carlo tree search, novel neighbourhood moves, memetic algorithms, and hyper-heuristic methods.
 The implementation is also engineered to increase the speed with which iterations are performed, and to exploit the computing power of multicore machines.
 Empirical evaluation shows that the resulting information-sharing multi-component algorithm significantly outperforms other solvers on a set of ``hidden'' instances, i.e.\ instances not available at the algorithm design phase.
\end{sloppypar}

\bigskip

\end{abstract} 

\maketitle

\section{Introduction}
\label{sec:intro}

\begin{sloppypar}
Project scheduling has been of long-standing interest to academics as well as practitioners. 
Solving such a problem requires scheduling of interrelated activities (jobs), potentially each using or sharing scarce resources, subject to a set of constraints, and with one or several of a variety of objective functions.
There are various project scheduling problems and many relevant surveys in the literature, e.g.\ see~\citep{Brucker1999,Herroelen1998,Herroelen2005,Linet1995,Moehring2003,Hartmann2010,Weglarz2011}. 
The best-known problem class is the Resource Constrained Project Scheduling Problem (RCPSP) in which activities have fixed usages of the resources, there are fixed precedence constraints between them, and often the objective is simple minimisation of the makespan (completion time of last activity). 
These problems have been proven to be NP-hard \citep{Blazewicz1983}, and a well-known benchmark suite,  PSPLIB, is provided in \cite{Kolisch1997}.
\end{sloppypar}

\begin{sloppypar}
A generalisation of the RCPSP is to also consider `Multi-mode RCPSP'  (MRCPSP) in which activities can be undertaken in one of a set of modes, with each mode potentially using different sets of resources.
Furthermore, there are many options besides makespan for the objective function(s); a typical one is that a weighted sum of completion times is minimised. 
As common in optimisation problems, exact methods perform best on smaller instances and on larger instances heuristics and metaheuristics become necessary. 
Recent works on the MRCPSP range from exact approaches, such as, MILP \citep{Kyriakidis2012}, 
and branch-and-bound \cite{Vanhoucke2001},
to metaheuristics, such as, 
differential evolution \citep{Damak:2009}, 
estimation of distribution algorithms \citep{Wang2012}, 
evolutionary algorithms \citep{Elloumi2010,Sebt2013,Gomes2014}, 
swarm intelligence methods \cite{KoulinasEtal2014:PSO-RCPSP}, 
and others \cite{ChenEtal2010:hybrid-RCPSP,WangFang2011:frog-MRCPSP}.
\end{sloppypar}

\begin{sloppypar}
This paper presents our winning approach submitted to MISTA 2013 challenge\footnote{\url{http://gent.cs.kuleuven.be/mista2013challenge/}} on a further extension called `multi-mode resource-constrained multi-project scheduling' (MRCMPSP) and the results on the associated benchmark/competition instances.
The full description of this problem domain can be found on the competition website and in \cite{Wauters2014}; however, for completeness we also summarise it in~Section~\ref{sec:pr}.
The broad aim is to schedule a set of different and partially interacting projects, with each project consisting of a set of activities. 
There are no precedence constraints between the activities of different projects however they can compete for resources. 
Also, the objective function is extended to be a mix of a kind of weighted completion time and makespan.
The MRCMPSP is hence interesting in that it has a mix of structures and requirements that are a step towards modelling the complexity of real-world scheduling problems. 
The high real-world relevance of the multi-project version of scheduling is well-known, e.g.\ a survey \citep{Lova2000} found that 
``84\% of the companies which responded to the survey indicated that they worked with multiple projects''.
However, the majority of scheduling work is on the single project version, though there is some existing work on the multi-project case, e.g.\ see~\citep{Lova2000,Goncalves2008,Krueger2009,Mendes2009}.
\end{sloppypar}

Our approach searches the space of sequences of activities, from which schedules are constructed and then the quality of each schedule is evaluated using the objective function.
The search process on the set of sequences operates in two phases in a ``construct and improve'' fashion.
In the first phase, a heuristic constructor creates initial sequences of activities. 
A novel proposal in this paper is to investigate the overall global structure of the solutions and use this to motivate constructing the initial sequences using a Monte-Carlo Tree Search (MCTS) method, e.g.\ see \citep{BrowneEtal2012:MCTS}.
This construction phase is followed by an improvement phase which makes use of a large and diverse set of heuristic neighbourhood moves. 
The search process during the improvement phase is carefully controlled by a combination of methods arising from a standard metaheuristic, namely memetic algorithm, and also an extension of existing hyper-heuristic components \citep{HysstPATAT,KheiriDom}. 

There is an interesting potential for dual views of the overall problem. 
It is defined as a multi-project problem,  but it can be also viewed as a single project (multi-mode) RCPSP, in which the precedence graph has a particular structure, consisting of disjoint clusters.
There is a sense in which we work with both views together.
Some neighbourhood moves treat the problem in a single-project fashion and work on the constituent activities; 
other neighbourhood operators explicitly consider the multi-project nature of the problem, and focus on moves of projects. 
Both views, and kinds of operators, are used and work together to improve the overall project-level structure as well as the detailed activity level structure.
A discussion and a computational study on both approaches can be found in~\citep{Lova2000}. 

The primary contributions of this paper are:
\begin{itemize}
	\item 
	Observation and investigation of how the primary objective function being essentially a ``sum of project completion times'' leads to good solutions having inherently different structure to those with makespan as the primary objective.  
	In particular, minimisation of project completion times subject to limited global resource results in partial ordering of projects; this does somewhat reduce the effective size of the search space, but also may lead to good solutions being more widely separated. 
	Understanding of this significantly affected our algorithm design, including an MCTS construction method aiming to create solutions having such structure.

	\item 
	Novel neighbourhood moves, including those that are designed specifically for smoother navigation through the search space of the multi-project extension of MRCPSP -- reflecting our observation of the effect that the main objective function has on the solution structures.

	\item 
	An adaptive hybrid hyper-heuristic system to effectively control the usage of the rich set of neighbourhood moves.

	\item 
	Evidence of the effectiveness based on successful results on a range of benchmark problems.
	This includes winning a competition, in which some problems were hidden at the algorithm design/tuning phase.
	We also tested our algorithm on single-project instances from PSPLIB\@.
	Although our algorithm was not designed to work on single-project instances, it demonstrated good performance in these tests, and was competitive with the state-of-the-art methods tailored to the single-project case.
	Furthermore, it improved 3 best solutions on these PSPLIB instances during these experiments.
\end{itemize}

These contributions are directed towards a system that is both robust and flexible; with the potential to be effective at handling a wide variety of problem requirements and instances.
Arguably, one of the lessons of this paper is that greater complexity and richness of such scheduling problems needs to be matched with a greater complexity and richness of the associated algorithms; especially when not all instances are known in advance, and so algorithms should not over-specialise to a particular data set.

Regarding the structure of the paper; in Section~\ref{sec:pr} we describe the problem to be solved.
In Section~\ref{sec:generator} we discuss how we have carefully chosen the appropriate data structures and implemented algorithms operating with those data structures efficiently in order to construct the schedule from a given sequence as fast as possible. (To build an effective system, one has to pay attention to all of its components.)
However, most of the contribution of this paper arises from choosing and combining the effective algorithmic components including the search control algorithm and then (partially) tuning the relevant parameters within the overall approach. 
These consist of the MCTS-based constructor given in Section~\ref{sec:constructor}, the neighbourhoods given in Section~\ref{sec:moves}, and the improvement phase given in Section~\ref{sec:imp}.
The computational experiments and competition results are presented and analysed in Section~\ref{sec:expt}; including some reports of performance on a multi-mode, though single project, benchmark set from PSPLIB.
Section~\ref{sec:conc} concludes the paper.

\myclearpage
\section{Problem Description}
\label{sec:pr}

\begin{sloppypar}
 The problem consists of a set $P$ of \emph{projects}, where each project $p \in P$ is composed of a set of \emph{activities}, denoted as $A_p$, a partition from all activities $A$.  
 Each project $p \in P$ has a \emph{release time} $e_p$, which is the earliest start time for the activities $A_p$.
\end{sloppypar}

The activities are interrelated by two different types of constraints: the \emph{precedence constraints}, which force each activity $j \in A$ to be scheduled to start no earlier than all the immediate predecessor activities in set $\mathit{Pred}(j)$ are completed; and the \emph{resource constraints}, in which the processing of the activities is subject to the availability of resources with limited capacities. 
There are three different types of the resources: local renewable, local non-renewable and global renewable.  Renewable resources (denoted using the superscript $\rho$) have a fixed capacity per time unit.  Non-renewable resources (denoted using the superscript $\nu$) have a fixed capacity for the whole project duration.  Global renewable resources are shared between all the projects while local resources are specified independently for each project.

 Renewable and non-renewable resources are denoted using the superscript $\rho$ and $\nu$, respectively.  
 $\mathscr{R}^{\rho}_p$ is the set of local renewable resources associated with a project $p \in P$, and $R_{pk}^{\rho}$ is the \emph{capacity} of $k \in \mathscr{R}^{\rho}_p$, i.e.\ the amount of the resource $k$ available at each time unit.  
 $\mathscr{R}^{\nu}_p$ is the set of local non-renewable resources associated with a project $p \in P$, and $R_{pk}^{\nu}$ is the capacity of $k \in \mathscr{R}^{\nu}_p$, i.e.\ the amount of the resource $k$ available for the whole duration of the project.  
 $\mathscr{G}^{\rho}$ is the set of the global renewable resources, and $G_k^{\rho}$ is the capacity of the resource $k \in \mathscr{G}^{\rho}$.

Each activity $j \in A_p$, $p \in P$, has a set of execution modes $\mathscr{M}_{j}$.  Each mode $m \in \mathscr{M}_j$ determines the duration of the activity $d_{jm}$ and the activity resource consumptions.  For a local renewable resource $k \in \mathscr{R}^{\rho}_p$, the resource consumption is $r^{\rho}_{jkm}$; for a local non-renewable resource $k \in \mathscr{R}^{\nu}_p$, the resource consumption is $r^{\nu}_{jkm}$; for a global renewable resources $k \in \mathscr{G}^{\rho}$, the resource consumption is $g^{\rho}_{jkm}$.

Schedule $D = (T, M)$ is a pair of time and mode vectors, each of size $n$.  For an activity $j \in A$, values $T_j$ and $M_j$ indicate the start time and the execution mode of $j$, respectively.  Schedule $D = (T, M)$ is feasible if:
\begin{itemize}
	\item For each $p \in P$ and each $j \in A_p$, the project release time is respected: $T_j \ge e_p$;
	\item For each project $p \in P$ and each local non-renewable resource $k \in \mathscr{R}^{\nu}_p$, the total resource consumption does not exceed its capacity $R_{pk}^{\nu}$.
	\item For each project $p \in P$, each time unit $t$ and each local renewable resource $k \in \mathscr{R}^{\rho}_p$, the total resource consumption at $t$ does not exceed the resource capacity $R_{pk}^{\rho}$.
	\item For each time unit $t$ and each global renewable resource $k \in \mathscr{G}^{\rho}_p$, the total resource consumption at $t$ does not exceed the resource capacity $G_k^{\rho}$.
	\item For each $j \in A$, the precedence constraints hold: $T_j \ge \max_{j' \in \mathit{Prec}(j)} T_{j'} + d_{j'M_{j'}}$.
\end{itemize}

\bigskip

The objective of the problem is to find a feasible schedule $D = (T, M)$ such that it minimises the so-called \emph{total project delay} (TPD), defined by
using the time for \emph{total project completion} (TPC)%
\begin{equation}
\label{eq:tpc}
\text{TPC} =  \sum_{p \in P} C_p 
\end{equation}
and
\begin{equation}
\label{eq:tpd}
\text{TPD} \equiv f_\text{d}(D) = \text{TPC} - L = \left( \sum_{p \in P} C_p \right) - L \,,
\end{equation}
where $C_p$ is the completion time of project $p$
\begin{equation}
\label{eq:compP}
C_p = \max_{j \in A_p} \left( T_j + d_{jM_j} \right) \,.
\end{equation}
The constant $L$ is a lower bound calculated as
\begin{equation}
L = \sum_{p \in P} \left( \text{CPD}_p + e_p \right) \,,
\end{equation}
with $\text{CPD}_p$ being a given pre-calculated value.  
Since $L$ is a constant, then it does not affect the optimisation
(it was presumably introduced in the competition just to make the output numbers smaller and easier to interpret).
Specifically, since $L$ is the lower bound (though not necessarily a tight bound), $f_\text{d}(D) \ge 0$ for any feasible solution $D$.

\begin{sloppypar}
Note that this primary objective is an instance of the standard ``weighted completion time'', usually denoted by ``$\sum_j w_j C_j$'', but specialised to the case, ``$\sum_p w_p C_p$'', in which only the completion times of the projects are used\footnote{If there is no unique activity marking the end of a project, then a dummy empty end activity can always be added, without changing the problem.} (in the case of TPD all the weights are assigned to be one).
\end{sloppypar}

The tie-breaking secondary objective is to minimise the \emph{total makespan}, (TMS), which is the finishing time of the last activity (or equivalently of the last project): 
\begin{equation}
\label{eq:f2}
\text{TMS} \equiv f_\text{m}(D) 
= \max_{j \in A} \left( T_j + d_{jM_j} \right) 
= \max_{p \in P} C_p \,.
\end{equation}

In our implementation, we combine the objective functions $f_\text{d}(D)$ and $f_\text{m}(D)$ into one function $f(D)$ that gives the necessary ranking to the solutions:
\begin{equation}
	\label{eq:f}
	f(D) = f_\text{d}(D) + \gamma f_\text{m}(D) \,,
\end{equation}
where $0 < \gamma \ll 1$ is a constant selected so that $\gamma f_\text{m}(D) < 1$ for any solution $D$ produced by the algorithm.  In fact, we sometimes use $\gamma = 0$ to disable the second objective.  For details, see Section~\ref{sec:memetic}. 

Under the conventional ``$\alpha\vert\beta\vert\gamma$'' labelling, it could perhaps be described as 
 ``MP$^+$S \textbar \, \textit{prec}  $\vert$ ( $\sum_p w_p C_p$, $C_\text{max}$ )''
using `MP$^+$S' to denote `Multi-mode Multi-Project Scheduling'.

\myclearpage
\section{Schedule Generator}
\label{sec:generator}

 Designing an algorithm for solving the multi-mode resource-constrained multi-project scheduling problem requires an appropriate solution representation.
There are two `natural' solution representations in the scientific literature:
\begin{description}
\item[Schedule-based:] A direct representation using the assignment times, and also modes, of activities, i.e.\ vectors $T$ and $M$.

\item[Sequence-based:] This is based on selecting a total order on all the activities.
Given such a sequence, a time schedule is constructed by taking the activities one at a time in the order of the sequence and placing each one at the earliest time slot such that feasibility of the solution would be preserved.  This approach is called \emph{serial schedule generation scheme} (e.g., see \cite{Brucker1999}).
\end{description}

The schedule-based representation is perhaps the most natural one for a mathematical programming approach, but we believe that it could make the search process difficult for a metaheuristic method, in particular, generating a feasible solution at each step could become more challenging.  
As is common for heuristic approaches \citep{Wauters2014}, we preferred the sequence-based representation, since it provides the ease of producing schedules that are both feasible and for which no activity can be moved to an earlier time without moving some other activities (the schedule is then said to be `active').

The sequence-based representation is a pair $S = (\pi, M)$, where $\pi$ is a permutation of all the activities $A$, and $M$ is a modes vector, same as in the direct representation.  
The permutation $\pi$ has to obey all the precedence relations, i.e., $\pi(j) > \pi(j')$ for each $j \in A$ and $j' \in \mathit{Pred}(j)$.  
The modes vector is feasible if $M_j \in \mathscr{M}_j$ for each $j \in A$ and the local non-renewable resource constraints are satisfied for each project $p \in P$.

In order to evaluate a solution $S$, it has to be converted into the direct representation $D$.  By definition, the sequence-based representation $S = (\pi, M)$ corresponds to a schedule produced by consecutive allocation of activities $\pi(1)$, $\pi(2)$, \ldots, $\pi(n)$ to the earliest available slot.  
The corresponding procedure, which we call \emph{schedule generator}, is formalised in Algorithms~\ref{alg:schedule-construction}, \ref{alg:scheduling-job} and \ref{alg:testing-slot}.  
Note that the procedure guarantees feasibility of the resulting schedule as it schedules every activity in such a way that feasibility of the whole schedule is preserved.

\begin{algorithm2e}[ht]
\caption{Serial schedule generation scheme}
\label{alg:schedule-construction}
	Let $S = (\pi, M)$ be the sequence-based solution\;
	\For {$i \gets 1, 2, \ldots, n$}
	{
		Let $j \gets \pi(i)$\;
		Schedule $j$ in mode $M_j$ to the earliest available slot such that feasibility of the schedule is preserved;
	}
\end{algorithm2e}

\SetKw{KwGoTo}{go to}


\begin{algorithm2e}[ht]
\caption{Scheduling an activity to the earliest available slot.}
\label{alg:scheduling-job}
	Let $j$ be the activity to be scheduled\;
	Let $m$ be the mode associated with $j$\;
	Let $p$ be an index such that $j \in A_p$\;
	Calculate the earliest start time of $j$ as $\displaystyle{t_0 \gets \max \left\{ e_p, \ \max_{j' \in \mathit{Prec}(j)} \left( T_{j'} + d_{j' m} \right) \right\}}$\;
	$t \gets \mathit{TestSlot}(j, t_0)$\;
	Allocate activity $j$ at $t$ in mode $m$ and update the remaining capacities\;
\end{algorithm2e}

\begin{algorithm2e}[ht]
\caption{A naive implementation of the $\mathit{TestSlot}(j, t)$ function.  The function returns the earliest time slot at or after $t$ feasible for scheduling activity $j$.}
\label{alg:testing-slot}
	Let $m$ be the mode associated with $j$\;
	Let $p$ be an index such that $j \in A_p$\;
	\For {$t' \gets t,\ t + 1,\ \ldots,\ t + d_{jm} - 1$}
	{
		\For {$k \in \mathscr{R}^{\rho}_p$}
		{
			Let $a$ be the remaining capacity of $k$ at $t'$\;
			\lIf {$r_{jkm}^{\rho} > a$}{\Return {$\mathit{TestSlot}(j, t + 1)$}}
		}
		\For {$k \in \mathscr{G}^{\rho}$}
		{
			Let $a$ be the remaining capacity of $k$ at $t'$\;
			\lIf {$r_{jkm}^{\rho} > a$}{\Return {$\mathit{TestSlot}(j, t + 1)$}}
		}
	}
\Return {$t$}\;
\end{algorithm2e}


The worst case time complexity of this implementation is $O(n (\zeta + T \rho d))$, where $\zeta = \max_{j \in A} |\mathit{Prec}(j)|$ is the maximum length of the precedence relation list, $T$ is the makespan, $\rho = |\mathscr{G}^{\rho}| + \max_{p \in P} |\mathscr{R}_p^{\rho}|$ is the maximum number of local and global renewable resources and $d = \max_{j \in A} d_{jM_j}$ is the maximum activity duration.  
The first term of the sum corresponds to handling precedence relations, and the second term corresponds to scanning slots and testing resource availability.  
 Note that $\zeta < n$, and the maximum number $\rho$ of resources is a constant in our benchmark instances.  
 Also, $T$ is typically linear in $n$, and, hence, the time complexity is quadratic.
 The schedule generator is the performance bottleneck of our solver (note that most of the local search moves described in this paper are no worse than linear time complexity).  
In our experiments, schedule generation was usually taking over 98\% of the CPU time.  
By introducing several improvements (based on information sharing and caching of partial solutions) described below we reduced the running times of the schedule generator by a factor of around ten compared to our initial routine implementation.
That significantly increased the number of iterations the higher level algorithm was able to run within a given time.

\myclearpage
\subsection{Issues in Efficient Implementation}
\label{sec:code-opt}

In this section, we briefly discuss algorithmic and implementational issues that do not directly affect the number of sequence evaluations, but that are designed to increase the rate that evaluations are performed.  
This is of importance for application of the methods to real-world problems.
However, it can also be of potential importance to choices between different heuristics or other algorithmic components. 

The methods for evaluation and comparison of algorithms are not necessarily clear.  In particular, two aspects that arise with respect to this work, and scheduling in general, are ``hidden instances'', and ``termination criteria''. Hidden instances (meaning ones that are not available until after the implementation is finished) act against the danger that occurs with open instances of the techniques becoming tailored to the specific instances. For reliability and verification of results this generally means the implementations must be finalised before the release of the instances. In practice, this seems to be rarely applied outside of the context of a competition. In such cases, one might regard competitions are a way to enforce the scientific good practice of fully finalising the algorithm and implementation before the testing. 

 The other aspect is evaluating algorithms' performance purely in terms of their (wall-clock) runtimes, or instead using some attempt at implementing independent ``counting'' measure of steps taken. 
 The advantages of the former ``runtime'' method is that it relates to what real world users would usually care about, and also might be the only real option when no sensible pure counting methods are available. 
 The advantage of the counting is that it hides the implementation efficiency and hence allows to compare ``pure algorithm designs''.  
 In some research situations the classes of algorithms are sufficiently similar for a counting based comparison being viable, and then may well be standard e.g.\ in genetic algorithms the number of fitness evaluations is commonly used.  

 Hence, the counting-based approach encourages/supports rapid explorations of ideas for new algorithms, however exclusive usage would discourage developing new practical methods of improving algorithm performance. 
 For example, counting fitness evaluations can miss the advantages of the incremental evaluation techniques routinely used in metaheuristics, and that are vital for their effectiveness.  
 Runtime-based approach on contrary encourages the researchers to exploit methods that are practical in real circumstances, taking into account incremental evaluation, parallelism and other considerations crucial for real-world systems.
 We note here that the associated added complexity of algorithm engineering could potentially be partly addressed by hyper-heuristics, as they could provide feedback to the programmer regarding the extra value of low-level heuristics if their implementation were improved.
 This is particularly relevant in the context of a solver employing multiple neighbourhoods, like the one presented in this paper.

 We believe that there is no simple answer to which of these two algorithm evaluation approaches is best in general and so both of them have their place.
 However in the context of the very well studied project scheduling we do believe that engineering questions need to be accounted for.
 As an example of the importance of such `engineering issues' we refer to another well-studied area of solving propositional satisfiability (SAT) problems and that has been active for many decades\footnote{E.g. competitions have been held for many years, see http://www.satcompetition.org/}. 
 An important part of development of SAT solvers was the development of `watched literals' \cite{MoskewiczEtal2001:zChaff}. 
The technique only directly affected the standard and routine `unit propagation' procedure in SAT solvers (which is the CPU-intensive portion, analogous to the schedule generator), but it did so in a fashion that meant new heuristics were then practical, leading to new algorithm designs.

In real-world usages of scheduling, an important aspect is the software engineering aspect of the time and cost of implementing and maintaining the software.
An initial implementation of a neighbourhood is often relatively easy; however, the practical problems can arise when effective use requires that it is implemented in a fashion that uses incremental or delta evaluation (so that the objective function does not require a full re-evaluation). 
To support the incremental evaluation it is often necessary to implement appropriate data structures that are more sophisticated, and so harder to implement and maintain.
With serial generation the majority of the CPU time is spent in the generation of the schedule from the sequence. 
Hence, we naturally found that significant improvements were achieved by modifying the generator algorithm.
Good `engineering' of the serial generation also has the important advantage that it helps all of the neighbourhoods. 
(If all the neighbourhoods were to rely on entirely separate implementations, then there would be much more pressure, for practical software engineering issues, to reduce to a smaller set.)

Observe (see Algorithm~\ref{alg:scheduling-job}) that the schedule generation algorithm spends most of the time finding the first available slot for an activity.  
To speed up this phase, we use a modification of the Knuth-Morris-Pratt substring search algorithm.  
By testing resource availability in the reversed order, we can use early exploration of insufficient resources to skip several values of $t$, see Algorithm~\ref{alg:scheduling-job-improved}.

\begin{algorithm2e}[ht]
\caption{An improved implementation of the $\mathit{TestSlot}(j, t)$ function.  The function returns the earliest time slot at or after $t$ feasible for scheduling activity $j$.}
\label{alg:scheduling-job-improved}
	\For {$t' \gets t + d_{jm} - 1,\ t + d_{jm} - 2,\ \ldots,\ t$}
	{
		\label{lab:test-t-improved}
		\For {$k \in \mathscr{R}^{\rho}_p$}
		{
			Let $a$ be the remaining capacity of $k$ at $t'$\;
			\lIf {$r_{jkm}^{\rho} > a$}{\Return {$\mathit{TestSlot}(j, t' + 1)$}}
		}
		\For {$k \in \mathscr{G}^{\rho}$}
		{
			Let $a$ be the remaining capacity of $k$ at $t'$\;
			\lIf {$r_{jkm}^{\rho} > a$}{\Return {$\mathit{TestSlot}(j, t' + 1)$}}
		}
	}
	\Return {$t$}
\end{algorithm2e}

 Another speed-up heuristic is exploiting the nature of the neighbourhood moves; we noted that any two solutions tested consequently are likely to share a prefix.  
 Let $S^1 = (\pi^1, M^1)$ be some solution, $S^2 = (\pi^2, M^2)$ be its neighbour, and $D^1 = (T^1, M^1)$ and $D^2 = (T^2, M^2)$ be their direct representations.  
 According to our assumption, $\pi^1(i) = \pi^2(i) = j$ and $M^1_j = M^2_j$ for $i = 1, 2, \ldots, x$, where $x$ is the prefix length (which is likely to be significant).  
 Then, by construction, $T^1_j = T^2_j$ for each $j = \pi^2(1),\linebreak[1] \pi^2(2),\linebreak[1] \ldots,\linebreak[1] \pi^2(x)$.  
 Hence, knowing $D^1$, we do not need to calculate the values $T^2_j$ for $j = \pi^2(1), \pi^2(2), \ldots, \pi^2(x)$.  
For details, see Algorithm~\ref{alg:prefix}.

 This gives a form of incremental evaluation that has the advantage that it applies to all the neighbourhoods used.
 Potentially, this engineering optimisation could have impact on the design of the improvement algorithm in that the selection of neighbourhood moves could benefit from exploiting changes at the end of the schedule being faster to evaluate than those at the beginning. 
 This is something that can potentially be captured by a hyper-heuristic, as one of the long-term intentions of hyper-heuristics is that they should monitor the CPU times taken by different moves, and combine this with monitoring of their effects, in order to give better adaptive control of the improvement phase.

\SetKw{KwAnd}{and}

\begin{algorithm2e}[ht]
\caption{Serial schedule generation scheme with prefix detection.}
\label{alg:prefix}
	Let $S^2 = (\pi^2, M^2)$ be the new solution\;
	Let $S^1 = (\pi^1, M^1)$ be the previous solution and $D^1 = (T^1, M^1)$ be the corresponding direct representation\;
	Let $\mathit{prefix} \gets \mathit{true}$\;
	\For {$i \gets 1, 2, \ldots, n$}
	{
		Let $j \gets \pi^2(i)$\;
		\eIf {$\mathit{prefix} = \mathit{true}$ \KwAnd $\pi^1(i) = j$ \KwAnd $M^1_j = M^2_j$}
		{
			Allocate activity $j$ to $T^1_j$ in mode $M^2_j$ and update the remaining capacities\;
		}
		{
			Schedule $j$ in mode $M^2_j$ to the earliest available slot\;
			$\mathit{prefix} \gets \mathit{false}$\;
		}
	}
\end{algorithm2e}



In addition to the algorithmic improvements, we used several standard programming techniques to optimise the implementation performance, with the primary ones being:
\begin{itemize}
	\item \textbf{Incremental maintenance of auxiliary data:} Note that the schedule generator has to maintain the amount of remaining resource for each renewable resource and each time unit, i.e.\ $T \cdot \big(|\mathscr{G}^{\rho}| + \sum_{p \in P} |\mathscr{R}^{\rho}_p|\big)$ values.
	Since $T$ is not known in advance, the corresponding auxiliary data structure has to be large enough to fit a schedule of size $T_\text{max}$, where $T_\text{max}$ is the upper bound of $T$.
    Considering the initialisation of this auxiliary data structure, the real time complexity of the algorithm is $O(n (\zeta + T \rho d) + T_\text{max} \rho')$, where $\rho' = |\mathscr{G}^{\rho}| + \sum_{p \in P} |\mathscr{R}_p^{\rho}| = O(\rho |P|)$.
	Usually $T_\text{max} \gg T$ and $\rho' \gg 1$ and, hence, the last term of the generator complexity has a major impact on the real performance of the procedure.
	By reusing the same auxiliary data structures and only re-initialising the portion that was altered in the previous run of the generator, we speed up the initialisation phase by a factor of about $T_\text{max} / T$.
	
	\ignore{
	To avoid dynamic memory reallocation, one has to allocate $U$ memory for each of such arrays, where $U$ is the upper bound of $T$.  Due to the lack of a good upper bound, $U$ is often much larger than $T$, and the initialisation phase of schedule generator, though linear, might take significant CPU time.
	
	Before each schedule generator run and instead of allocating and initialising $(|\mathscr{G}^{\rho}| + \sum_{p \in P} |\mathscr{R}^{\rho}_p|) U$ memory, we reuse the previously allocated memory.  Let $D'$ be the previous schedule produced by the generator.  To evaluate it, we calculate its exact makespan $U' = f_\text{d}(D')$ (see~(\ref{eq:f2})).  Now, to make sure that the resource availability arrays are initialised correctly, it is enough to reinitialise the first $U'$ elements of each of them.
	
}
	
	\item \textbf{Local memory access:} 
	Most of the solution evaluations happen in our metaheuristic within local search procedures.  Each instance of the local search algorithm is assigned a CPU core, and a dedicated copy of the schedule generator is maintained for it.  This is particularly important for the prefix reuse heuristic and CPU cache efficiency.

\ignore{
\AJP{This is a bit too messy, and detailed.}
	\item We noticed that most of the test problems have exactly one global renewable resource and one local renewable resource.  Several more problems have two global renewable and no local renewable resource.  We implemented corresponding special cases of the schedule generator, which had a significant impact of the running time of the algorithm.  Indeed, in case of one global and one local resources, we were able to notably simplify data structures and exclude one of the nested loops.  In case of none local resources, we excluded all the logic related to the local resources.
	
}

\end{itemize}

\myclearpage
\section{Solution Structures}
\label{sec:constructor}

This section primarily explains why having ``Total Project Delay'' (TPD) as the main objective can be expected to lead to the general structure of solutions being very different to the structure obtained with the more standard makespan (TMS) objective.
Our observations of the TPD-driven structure had an important influence on our design of the overall algorithm. 
The structure motivates many of the ``project level'' moves that are considered in Section~\ref{sec:moves}. 
In this section, we also propose a heuristic method to construct the initial solutions which feed into the improvement phase later. 
As standard in such ``construct and improve'' optimisation, the concern was that the desirable structures might be difficult to achieve by the improvement unless the initial constructor heuristically tried to get close, by doing an appropriate ``heuristically-guided global sampling'' of the space of approximate structures.
Accordingly, with the intent to increase the robustness of the solver, to handle such structures in unseen instances, a specialised MCTS-based constructor is proposed and developed. 


\subsection{TPD-driven Solution Structures}
\label{sec:constructor:structure}

\newcommand{\tpdexample}[2]{%
  	\begin{tikzpicture}[scale=0.28]	
		\foreach \i in {1,...,2}{
			\expandafter\xdef\csname lastres@array@\i\endcsname{0}
			\expandafter\xdef\csname lastproj@array@\i\endcsname{0}
		}
		
		\foreach \i in {1,...,2}{
			\node[font=\small,anchor=east] at (-0.5, 3 - \i) {Res.\ \i};
		}

		\foreach \project/\duration/\resource/\lbl in {#1}{			
			\pgfmathparse{\csname lastres@array@\resource\endcsname}
			\let\leftres=\pgfmathresult

			\pgfmathparse{\csname lastproj@array@\project\endcsname}
			\let\leftproj=\pgfmathresult
			
			\pgfmathparse{max(\leftres, \leftproj)}
			\let\left=\pgfmathresult

			\pgfmathparse{\left + \duration}
			\let\right=\pgfmathresult;
			\expandafter\xdef\csname lastres@array@\resource\endcsname{\right};
			\expandafter\xdef\csname lastproj@array@\project\endcsname{\right};
		
			\expandafter\xdef\csname tms\endcsname{\pgfmathresult}

			\draw[job \project] (\left, \resource - 0.1) rectangle (\right, \resource + 0.1);

			\pgfmathparse{(\left + \right) * 0.5}
			\node[font=\footnotesize, anchor=south] at (\pgfmathresult, \resource) {$\lbl$};
		}

		\expandafter\xdef\csname tms\endcsname{0}
		\expandafter\xdef\csname tpd\endcsname{0}
		\foreach \i in {1,...,2}{
			\pgfmathparse{\csname tpd\endcsname + \csname lastproj@array@\i\endcsname}
			\expandafter\xdef\csname tpd\endcsname{\pgfmathresult}

			\pgfmathparse{max(\csname tms\endcsname, \csname lastproj@array@\i\endcsname)}
			\expandafter\xdef\csname tms\endcsname{\pgfmathresult}
		}
		
		\node[anchor=north west] at (0, 0) {#2};
	\end{tikzpicture}
}

\begin{figure*}[htb]
\begin{subfigure}[t]{1\textwidth}
	\includegraphics{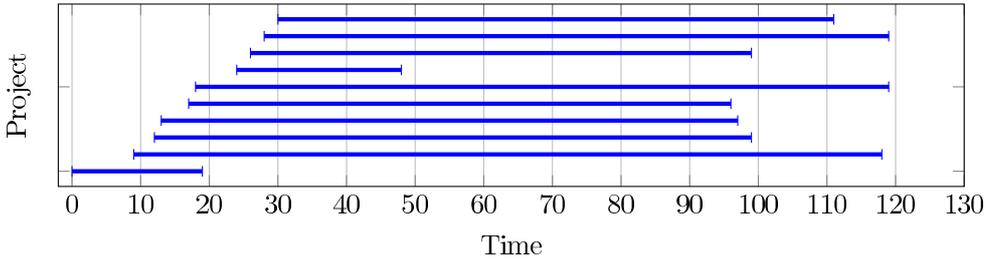}	
	\caption{
		Only makespan is minimised (the TPD objective is disabled), that is minimising the latest completion times of projects.
		$\text{TPC} = 925$, $\text{TMS} = 119$.}
	\label{fig:project-ordering-makespan}
\end{subfigure}
\\[4ex]
\begin{subfigure}[t]{1\textwidth}
	\includegraphics{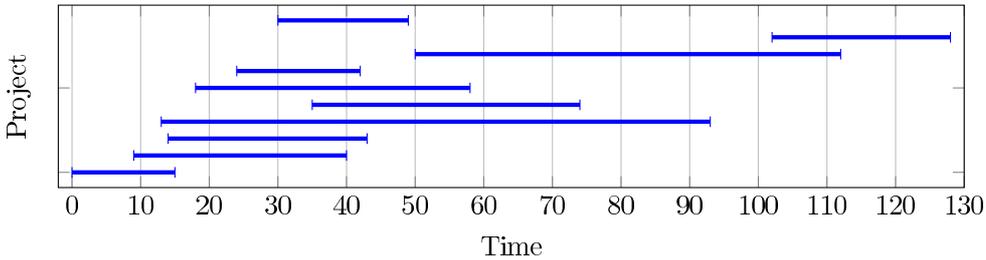}	
	\caption{
		Both objectives are enabled, that is, the primary objective is the TPD -- minimising the sum of completion times of projects.
		$\text{TPC} = 654$, $\text{TMS} = 128$.}	
	\label{fig:project-ordering-tpd}
\end{subfigure}
\caption{
	An example of overall project-level structure of good solutions (using instance B-1).
	Each horizontal line shows the duration allocated to each project in the schedule.}
\label{fig:project-ordering}
\end{figure*}

In this subsection, we report on the investigation of the natural question of what constitutes a `good' approximate structure.
We firstly look at the structure of a high quality schedule for a benchmark instance, and then we elucidate the observed structures using some small examples.
The critical message is that optimising TPD leads to different structures of the solutions than when optimising TMS; we believe that algorithm design needs to take account of this difference.\footnote{Similar effects are also observed in \cite{Wauters2015}.}

In particular, we have observed that in many  cases, dominance by the TPD objective frequently leads to approximate ordering of the projects.
A typical example of this is given in Figure~\ref{fig:project-ordering} using a good solution to instance B-1\@. 
Figure~\ref{fig:project-ordering-makespan} shows the structure obtained when only the standard makespan is minimised, but the structure is very different in  Figure~\ref{fig:project-ordering-tpd} with the required TPD-dominated objective. 
That is, the pattern of completion times of each project, the $C_p$ of (\ref{eq:compP}), depends on whether they are driven by TPD, effectively minimising the average of the $C_p$, or instead driven by the makespan, reducing the maximum of the $C_p$. 
The example shows how, with TPD dominating, there are time periods in the schedule when the general focus is on relatively few projects and during the schedule this focus changes between projects.

Hence, the evidence from Figure~\ref{fig:project-ordering}, and other multi-project cases we have looked at, suggests that the approximate ordering is common.
Overall, these structures naturally arises from the combination of the TPD (\ref{eq:tpd}) objective with the limited global resources.
Suppose that some P is the final finishing project, and so its last activity determines the TMS and P' is an earlier finishing project.
It may well be that P' can move some of its activities earlier by delaying activities of P, though without changing the last activity of P. 
Such moves will improve the TPD without worsening the TMS. 

That is, in contrast to the makespan objective, the TPD objective, together with limited shared resources, has a natural side-effect of encouraging unfairness between the finish times of projects, and will drive some projects to finish as early as possible.

\pgfplotsset{compat=newest}

\begin{figure*}[htb]

\begin{minipage}[c][11cm][t]{0.5\textwidth}
  \vspace*{\fill}
\begin{subfigure}{\textwidth}

\tikzset{
	job/.style={rectangle},
	job 1/.style={job, fill=black},
	job 2/.style={job, fill=green}
}

\begin{center}
\includegraphics{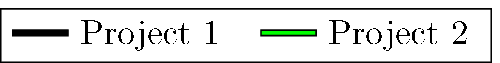}
\end{center}

\includegraphics{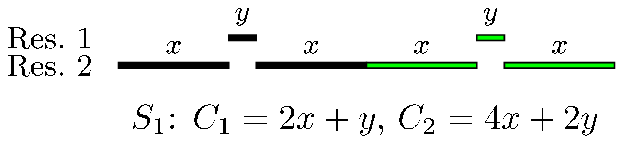}
\\[3ex]
\includegraphics{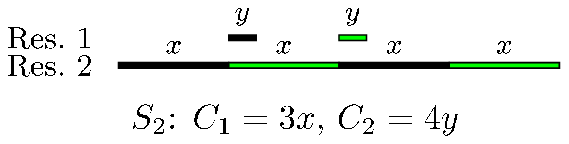}
\\[3ex]
\includegraphics{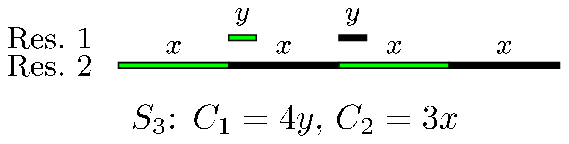}
\\[3ex]
\includegraphics{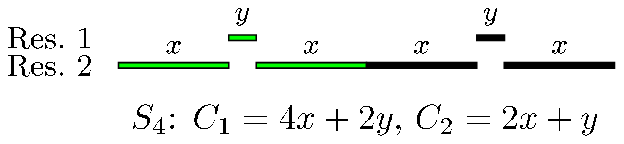}
	\caption{
		Case ``2[$x$--$y$--$x$]''. 
		Showing four possible solutions, $S_1$, \ldots, $S_4$, and associated project completion times $C_1$ and $C_2$.}
	\label{fig:TPD:solns}
\end{subfigure}
\end{minipage}%
\hspace{0.05\textwidth}
\begin{minipage}[c][11cm][t]{0.45\textwidth}
\vspace*{\fill}
\begin{subfigure}{\textwidth}  
\includegraphics{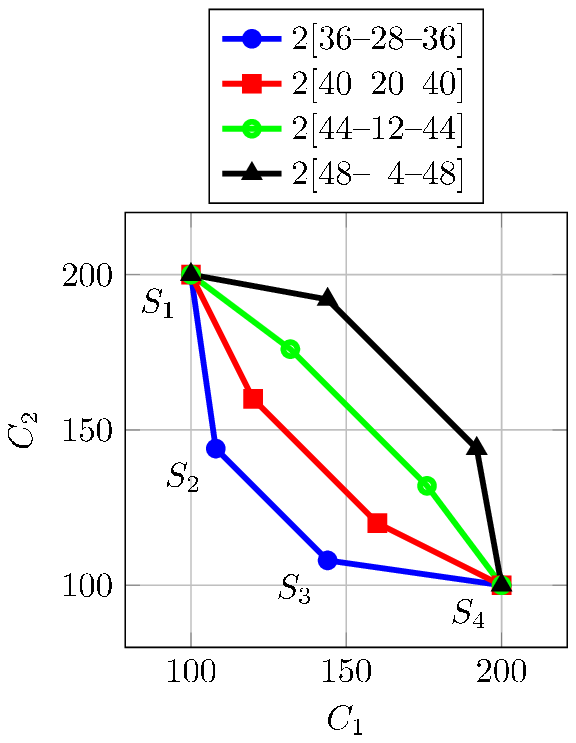}
\caption{Tradeoff between completion times of projects. We select a range of different values for $x$ and $y$, but with the invariant that $2x + y = 100$; so that the effect of varying the balance between $x$ and $y$ is made clear.}
\label{fig:TPD:tradeoff}
\end{subfigure}
\end{minipage}

\caption{
	Simple 2-project examples to illustrate concave and convex non-dominated sets.
	The notation ``2[$x$--$y$--$x$]'' means two projects each consisting of a chain of 3 tasks of durations $[x,y,x]$ with shared resources, and the tasks 1 and 3 in each chain using resource 2, and task 2 using resource 1.}
\label{fig:TPD}
\end{figure*}

To illustrate the way in which an approximate ordering might arise, and TPD-driven and TMS-driven solutions can be quite different, we give a set of small example instances.
We denote these example instances by ``2[$x$--$y$--$x$]'', consisting of two projects.
Each project consists of a precedence-constrained chain of just 3 activities of durations $x$, $y$ and $x$, respectively. 
There are two (renewable) shared resources with activities 1 and 3 using the resource 2, and activity 2 using resource 1.
We focus on values of $x$ and $y$ that lead to resource 2 being the bottleneck, and so the main driving force of the makespan -- corresponding to the shared global resource in the MISTA benchmarks.
Figure~\ref{fig:TPD:solns} shows four schedules of this parametrised instance, along with their corresponding values of $C_1$ and $C_2$.
Solutions $S_4$ and $S_3$ arise from simply swapping the order of the projects in solutions $S_1$ and $S_2$ respectively.

The solutions $S_1$ and $S_4$ do not interleave the two projects, and so leave gaps in the bottleneck resource 2. 
In contrast, solutions $S_2$ and $S_3$ interleave the projects and so lead to full usage of the bottleneck resource, and are hence automatically the best solutions for the TMS objective.
However, $S_2$ and $S_3$ are not always the best solutions for the TPD objective. 
Changing the solution leads to a tradeoff between $C_1$ and $C_2$, and this is illustrated, in Figure~\ref{fig:TPD:tradeoff} for various choices of $x$ and $y$.
In this figure, the values of $x$ and $y$ are selected so that we always have $2x + y = 100$; this makes the effect of changing the relative balance of $x$ and $y$ clearer, and also means that the values can be interpreted as percentages of the project time.

Before proceeding further, we observe that we can also interpret the project completion times, $C_p$, from the point of view of `multi-objective optimisation', by regarding each of them as separate objectives. 
In particular, in standard fashion, one can say that a set of values for $C_p$ dominates another set provided that none of the $C_p$ values are worse (larger) and at least one is better (smaller).
One can hence discuss the cases in  Figure~\ref{fig:TPD:tradeoff} as being the $C_p$-Pareto Front, of non-dominated sets of $C_p$.
We emphasise this view is for obtaining insight into the space of solutions, and is not used directly within our algorithm; we did not perform multi-objective optimisation over the $C_p$, though believe it would be worthy of future investigation. 
Also, note this view is different from taking the pair $(\text{TPD}, \text{TMS})$ as a bi-objective problem, and having a $(\text{TPD}, \text{TMS})$-Pareto Front (this would correspond again to a different class of algorithms than we consider here, but is again worthy of future investigation).
Incidentally, there are two distinct `multi-objective views' on the space of solutions:
\begin{itemize}
	\item  
	$C_p$-Pareto Front, of dimension $p$, formed from the $p$ completion times
	
	\item  
	$(\text{TPD}, \text{TMS})$-Pareto Front, of dimension 2, and formed from the aggregates (sum and max) of the $C_p$
\end{itemize}
However, it is important to realise that although the $C_p$ and $(\text{TPD}, \text{TMS})$ views are different, they are tightly linked.
Both the TPD, arising from $\sum_p C_p$,  and the TMS,  from $\max_p C_p$, are monotone (non-decreasing) with respect to the $C_p$ and so preserve Pareto dominance.
That is, if solution $C_p$ dominates solution $C'_p$, then the resulting pair $(\text{TPD},\text{TMS})$ dominates $(\text{TPD'},\text{TMS'})$ -- though note that the converse does not apply.
Hence, the $(\text{TPD}, \text{TMS})$-Front can be extracted from the $C_p$-Front.  
However, this `projection' loses information; the $C_p$-Front gives more insight into the space of solutions.

Consequently, the ``$C_p$-Pareto Front'', or set of non-dominated $C_p$ values, can give some insight into the effect of TPD compared to TMS on the structure of the solutions.
Hence, we can regard Figure~\ref{fig:TPD:tradeoff} as showing the effect of the $x$--$y$ balance on the $C_p$-Pareto Front.
In particular, we see that for two of the cases, 2[44--12--44] and 2[48--4--48] the $C_p$-Pareto Front is concave in terms of the set of feasible $C_p$ values (on the top right).
When the Front is concave then the two solutions minimising the TPD are $S_1$ and $S_4$ and are at opposite ends of the $C_p$-Front.
In contrast, the solutions $S_2$ and $S_3$ minimising the TMS are in the middle.

Generally, with a concave $C_p$-Pareto Front the solutions (locally) minimising $\sum_p C_p$ are naturally at the ends of the Front, and furthermore solutions at the ends of the Front will be more likely to have some $C_p$ values small and others large.
This matches with the way that the observed solutions minimising TPD do indeed sequence the projects. 
That is, when the projects are roughly equal size, and the $C_p$ Pareto Front is concave, then it becomes reasonable that the TPD objective prefers the end points, and so the candidate good solutions are more likely to be widely separated in the search space.

Also, notice that as the fraction of activities using the bottleneck resource 2 (that is the value of $2x$) increases, then the $C_p$-Pareto Front becomes more concave, and the TPD-driven solutions become different from the TMS-driven ones. 
This is consistent with the our general experience of the properties of MISTA instances; the shared global resource tends to be a bottleneck, and consequently they show the approximate ordering as seen in Figure~\ref{fig:project-ordering}.
Future work could well use such features related to the tightness of bottleneck resources to predict these effects, and so be used to select appropriate algorithm components. 
We also remark that the interesting structure of the tradeoff between the completion times does suggest that future work might well use multi-objective methods, e.g.\ see \cite{Ceollo99:MOO-survey}, and of course could study the effect of different weights for the completion times.

The simple example above is atypical in that the two projects are identical. 
However, if the projects are similar in size then it is reasonable to expect that the $C_p$-Pareto Front may be similar in overall structure but slightly distorted.
When the front is concave but still roughly symmetric with respect to swapping the $C_p$, then the TPD-driven solutions are likely to still be towards the ends of the Front and so widely separated.
However, because the front is not totally symmetric the ends are likely to have slightly different values of TPD. 
Hence, in such circumstances, it is reasonable to expect that there may be many widely separated local optima in TPD, and also that the optima correspond to different (approximate) orderings of the projects.

\subsection{MCTS} 
\label{sec:constructor:MCTS}

We have seen above that the TPD objective tends to favour a different solution structure than standard makespan minimisation, and in particular can drive a partial orderings of the projects, and there may be many good solutions that are widely separated in the solution space.
Consequently, our construction method attempts to do a broad exploration of the solution space and to create initial sequences with a broadly similar partial order.
The reason for including this was simply that we expected the partial ordering to lead to widely separated local minima in the search space, with large barriers between them -- as swapping the order of two projects might mean going through interleaved intermediate states in which the TPD is worse. 
We included a method to sample the space of approximate project orderings, with the intent to avoid starting in a poor ordering and being trapped.

We expect that only a partial ordering is needed because we can assume that the subsequent improvement phase can make small or medium size adjustments to the overall project ordering structure.
However, the improvement phase could have more difficulty, and take more iterations, if the general structure of the project ordering were not close to the structure expected in best solutions.
Consequently, and for simplicity, we decided that a reasonable approximation would be to use a 3-way partition of the projects taken to correspond to `start', `middle' and `end' parts of the overall project time. 
We required the numbers of projects in each part (of the partition) to be equal -- or with a difference of at most one when the total number of projects is not a multiple of 3.

The problem then is how to quickly select good partition of the projects, and the method we selected is a version of Monte-Carlo Tree Search (MCTS) methods \cite{BrowneEtal2012:MCTS}.
The general idea of MCTS is to search a tree of possibilities, but the evaluation of leaves is not done using a predefined heuristic, but instead by sampling the space of associated solutions.
The sampling is performed using multiple invocations of a ``rollout'' which is designed to be fast and unbiased.
It needs to be fast so that multiple samples can be taken; also rather than trying to produce ``best solutions'' it is usually designed to be unbiased -- the (now standard) idea being that it should provide reliable branching decisions in the tree, but is not directly trying to find good solutions.

In our case, the tree search corresponds to decisions about which projects should be placed in which part of the partition.
The rollout is a fast way to sample the feasible activity sequences consistent with the candidate choice for the partition of the projects.
Specifically, the tree search works in two levels; firstly to select the projects to be placed in the end part and then to select the partition between the start and middle parts.

The first stage considers\footnote{The parameters for the MCTS were the result of some mild tuning, and used in the competition submission, but of course are adjustable.}  100 random choices for the partition of the projects, and then selects between these using 120 samples or the rollout\footnote{The number 120 was selected so that the rollouts could be evenly distributed between 2, 4, 6, 8 or 12 cores of the machine}.  The rollout consists of two main stages:
\begin{enumerate}
\item Randomly select a total ordering of the activities consistent with the precedences and with the candidate partitioning. Specifically, within each partition we effectively consider a dispatch policy that randomly selects between activities that are available to be scheduled because their preceding activities (if any) are already scheduled.
\item Randomly select modes for the activities. If the result is not feasible then this can only be because of the mode selection causing a shortfall in some non-renewable resources. Hence, it is repaired using a local search on the space of mode selections.  We use moves that randomly flip one mode at a time, and an objective function that measures the degree of infeasibility by the shortfall in resources. Since the non-renewable resources are not shared between projects, this search turned out to be fast and reliable.  As a measure of precaution, if the procedure fails to obtain a feasible selection of modes after a certain number of local search iterations, we restart it with a random selection of modes.
\end{enumerate}
The first stage ends by making a selection of the best partitioning, using the quality of the 25th percentile of the final solution qualities (the best quartile) of the results of the rollouts.
The `end' part is then fixed to that of the best partition.
The decision to fix the `end' part also arose out of the observation that in good solutions the end projects are least interleaved.
The MCTS proceeds to the second stage, and follows the same rollout procedure but this time to select the contents of the middle (and hence start) parts.
This entire process usually completes within only a few seconds, and was used as the construction stage before the much longer improvement phase. 
We emphasise that the TPD-structure observed earlier had an important influence of the design of the neighbourhoods used in the key improvement phase -- in particular, the motivation that some moves should affect the project level structure.

\myclearpage
\section{Neighbourhood Operators}
\label{sec:moves}

%

In this section, we describe our neighbourhood operators, used by several components during the improvement phase.
The operators (also referred to as low-level heuristics or simply moves, depending on the algorithm which makes use of them) are categorised into three groups.
This categorisation is mainly based on the common nature of the strategy the operators employ while manipulating the solution. 
Some of the moves are similar to those used by other submissions; see \cite{Wauters2014}, and \cite{Geiger13} and \cite{toffolo2014} for few examples.
To make the paper self-contained, below we provide brief descriptions of all the moves used in our algorithm.

We guarantee that all of our operators preserve feasibility of the solution.
Also, all of the moves are randomised so that they could be repeatedly used in ``simulated annealing''-like improvement or applied as mutation operators.
All the random selections are made at uniform unless specified otherwise.

\def\move#1{{\sc #1}}

\subsection{Activity-level Operators}
\label{sec:moves-activity-level}

 Operators in this category involve basic operations, widely used in the literature, such as changing the mode of a single activity or swapping the positions of two activities.
 On top of that, we implemented a limited first improvement local search procedure for each of the basic moves.

 To describe the moves in this category, we will need additional notations.
 Let $\mathit{pos}(j)$ be the position of activity $j \in A$ within a given solution. 
 If activity $j$ is shifted to a different position, we say that the feasible range of its new position is $[\ell(j), u(j)]$, where $\ell(j)$ and $u(j)$ can be computed as follows:
 \begin{equation}
	\label{eq:insert-lower-bound}
		\ell(j_1) = \begin{cases}
			\max_{j \in \textit{Pred}(j_1)} \textit{pos}(j) + 1 & \text{if }\textit{Pred}(j_1) \neq \emptyset, \\
			1 & \text{otherwise}
		\end{cases}
 \end{equation}
 and
 \begin{equation}
	\label{eq:insert-upper-bound}
		\ell(j_1) = \begin{cases}
			\min_{j \in \textit{Succ}(j_1)} \textit{pos}(j) - 1 & \text{if }\textit{Succ}(j_1) \neq \emptyset, \\
			n & \text{otherwise,}
		\end{cases}
 \end{equation}
 where $\textit{Succ}(j_1)$ is the set of successors of $j_1$.
 
\begin{itemize}
	\item 
	\move{Swap activities}: swap two activities in the sequence.
	Select an activity $j_1 \in A$ randomly.
	Select another activity $j_2 \neq j_1 \in A$ randomly such that $\ell(j_1) \le \mathit{pos}(j_2) \le u(j_1)$.
	If $\ell(j_2) \le \mathit{pos}(j_1) \le u(j_2)$ then swap $j_1$ and $j_2$.
	Otherwise leave the solution intact.

	\item
	\move{Shift}: shift an activity to a new location in the sequence.
	Shift a randomly selected activity $j \in A$ to a new position randomly selected from $\mathit{pos}(j) \in [\ell(j), u(j)]$.

	\item
	\move{Change mode}: change the mode of a single activity.
	Select an activity $j$ randomly at uniform, and if $|\mathscr{M}_j| = 1$ then leave the solution intact.
    Otherwise select a new mode $m \neq M_j \in \mathscr{M}_j$ and, if this does not lead to a violation of non-renewable resource constraints, update $M_j$ to $m$.

	\item
	\move{FILS swap activities}: apply the first improvement local search (FILS) procedure based on the swap move.
	The operator has one parameter: the width $W > 1$ of the window to be scanned.
	Select an activity $j_1 \in A$ randomly.
	Define the window $[\ell', u']$ for activity $j_2$ as follows.
	If $u(j) - \ell(j) < W$, let $\ell' = \ell(j)$ and $u' = u(j)$.
	Otherwise select $\ell' \in [\ell(j), u(j) - W + 1]$ and set $u' = \ell' + W - 1$.
	Then, for every $j_2 \in A \setminus \{ j_1 \}$ such that $\ell' \le \mathit{pos}(j_2) \le u$, attempt to swap $j_1$ and $j_2$ (see \move{SwapActivities}).
	If the attempt is successful and it reduces the objective value of the solution then accept it and stop the search.
	Otherwise roll back the move and proceed to the next $j_2$ if any.

	\item
	\move{FILS shift}: apply the first improvement local search (FILS) procedure based on the shift move.
	The operator is implemented very similarly to \move{FILS swapActivities}, i.e.\ it attempts to shift a randomly selected activity to an new position within a window of a given size.

	\item
	\move{FILS change mode}: apply the first improvement local search (FILS) procedure based on the change mode move (the operator has no parameters).
	Select an activity $j \in A$ randomly.
	For $m \in \mathscr{M}_j \setminus \{ M_j \}$, produce a new solution by setting $M_j = m$.
	If the resulting solution is feasible and provides an improvement over the original solution, accept the move and stop the local search.
	Otherwise proceed to the next $m \in \mathscr{M}_j \setminus \{ M_j \}$ if any.
	
	\AJPii{Note this has some similarities to the mode repair procedure in \cite{PeteghemVanhoucke2011:mRCPSP-GA}}
	
\end{itemize}

\subsection{Ruin \&{} Recreate Operators}
\label{sec:moves-rnr}

 The Ruin \&{} Recreate (R\&{}R) operators are widely used in metaheuristics as mutations or strong local search moves but, to the best of our knowledge, they are relatively new to serial generation in scheduling. 
 As the name suggests, such operators have two phases: the ``ruin'' phase removes some elements of the solution and the ``create'' phase reshuffles those elements (for example, randomly) and then inserts them back.
 The number of elements to remove and re-insert is a parameter of a R\&{}R operator, which controls its strength (i.e.\ the average distance between the original and the resulting solutions).
 
 We implemented three different types of R\&{}R operators, and several strategies to select activities involved in the move.
 One can arbitrary combine any type of the R\&{}R operator with any activity selection strategy.
 
 The move types are \move{Reshuffle Positions}, \move{Reshuffle modes} and \move{Reshuffle positions and modes}.
 \move{Reshuffle positions} move removes all the selected activities $A' \subset A$ from the solution (leaving $|A'|$ gaps in the sequence) and then re-inserts them in a random order while respecting the precedence relations.
 To determine a feasible order, we compute which activities $V^i$ can be placed in each gap $i = 1, 2, \ldots, |A'|$ (in terms of precedence relations between the activities in $A'$ and activities in $A \setminus A'$), and also produce a precedence relations sub-graph induced by $A'$.
 Then we apply a backtracking algorithm.
 In each iteration, it fills one gap, starting from the earliest gaps in the sequence.
 For gap $i$ it randomly selects an activity $j \in A'$ such that there are no incoming arcs to $j$ in the precedence sub-graph.
 If such an activity exists, the algorithm removes $j$ from $A'$ and from the precedence sub-graph.
 Otherwise it rolls back to attempt another activity on the previous level of search.
 The depth of roll backs is unlimited, i.e., in the worst case, the algorithm will performs the depth first search of the whole search tree (observe that there is always at least one feasible arrangement of the activities).
 
 \move{Reshuffle modes} move changes the modes of selected activities while keeping their positions intact.
 Each of the selected activities $j \in A'$ is assigned a randomly chosen mode, either equal or not to the previous mode $M_j$.
 Observe that the new mode selection may cause infeasibility in terms of non-renewable resources.
 Thus, we use the multi-start metaheuristic, simply repeating the above step until a feasible mode selection is found.
 
 \move{Reshuffle positions and modes} combines the above two moves, which is trivial to implement as modes feasibility is entirely independent of the sequence feasibility.

 We implemented several activity selection strategies some of which exploit our knowledge of the problem structure.

\begin{itemize}
	\item
	\move{Uniform}: as the name suggests, $A' \subset A$ is selected randomly at uniform.
	The number of elements in $A'$ is a parameter of the move.

	\item
	\move{Project}: the activities $A'$ are selected within a single project, i.e.\ $A' \subset A_p$.
	The project $p \in P$ is selected randomly.
	
	\item
	\move{Local}: the selection of activities is biased to those scheduled near a certain time slot in the direct representation $D$.
	We randomly sample the activities accepting an activity $j$ with probability
	\begin{equation}
		\label{eq:move-local}
		\mathit{probability}(j) = \frac{1}{\frac{|T_j - \tau|}{\mathit{width}} + 1} \,,
	\end{equation}
	where $0 \le \tau \le f_\text{m}(D)$ (see (\ref{eq:f2})) is the randomly selected time slot (the centre of the distribution) and $\mathit{width}$ is a parameter defining the spread of the distribution.
	The sampling stops when $A'$ reaches the prescribed cardinality.

	\item 
	\move{Global resource driven}: the selection of activities is biased to the ones scheduled to time slots that under-utilise the global resources.
	The rationale is that global resources usually present a bottleneck in minimising the project completion times, and inefficiencies in the global resources consumption should be addressed when polishing the solution.
	In this selection strategy, as in \move{Local selection}, we use random sampling, but the acceptance probability for an activity $j \in A$ is defined by
	\begin{equation}
		\mathit{probability}(j) = \frac{\sum_{k \in \mathscr{G}^\rho} \mathit{remaining}_k(T_j)}{\sum_{k \in \mathscr{G}^\rho} G^\rho_k} \,,
	\end{equation}
	where $\mathit{remaining}_k(t)$ is the remaining (unutilised) capacity of global resource $k$ at the time slot $t$.
%
%
%

	\item 
	\move{Ending biased}: the selection of activities is biased to the last activities within the projects.
	The rationale is that, in an unpolished solution, the completion of a project might be improved by careful packing the last activities such that all the last activities end at roughly the same time, effectively maximising the utilisation of the local resources and redistributing the global resources between projects.
	As in \move{Local selection} and \move{Global resource driven selection}, we use random sampling of activities.
	The probability of accepting an activity $j \in A$ is
	\begin{equation}
		\mathit{probability}(j) = \frac{\varname{project-pos}(j)}{|A_p|} \,,
	\end{equation}
	where $p$ is the project of activity $j$ and $\varname{project-pos}(j)$ is the position of activity $j$ among the activities of project $p$.
%

\end{itemize}

\subsection{Project-level Operators}
\label{sec:moves-project-level}

 It was shown in Section~\ref{sec:constructor} that the TPD objective function tends to create a partial ordering of projects in high-quality solutions.
 The danger, however, is that the ordering to which the solution converges might be sub-optimal.
 Consider two well-polished solutions $S_1$ and $S_2$ having different project orderings.
 The distance between $S_1$ and $S_2$ is likely to be significant with respect to the activity-level and R\&{}R operators as many activities need to be moved to convert $S_1$ into $S_2$.
 Moreover, the transitional solutions in such a conversion will have significantly poorer quality compared to that of $S_1$ and $S_2$.
 This indicates that, once the algorithm converges to a certain project ordering, it needs a lot of effort to leave the corresponding local minimum and change the project ordering.
 
 The project-level operators are designed to overcome such barriers, allowing faster exploration of the rough landscape of the MRCMPSP\@.
 They perform on the project level and thus one move of any of operators from this category usually results into a change in the project ordering.
 That being said, the project-level moves are likely to corrupt the solution causing many activity-level inefficiencies that need to be treated with activity-level and R\&{}R moves.
 Thus, they ought to be used rarely within local search, but they can serve well as mutation operators.
 
 In some of the project-level moves we use the formal concept of project ordering $P(S)$.
 To extract $P(S)$ from a solution $S$, we compute the ``centre of mass'' $\mathit{centre}(p)$ for each project $p \in P$ as
 $$
 \mathit{centre}(p) = \frac{1}{|A_p|} \sum_{j \in A_p} \mathit{pos}(j)
 $$
 and define the ordered set $P(S)$ of projects according to their centres of mass.
 
\begin{itemize}
	\item
	\move{Swap two projects}: swap two randomly selected projects in the sequence.
	Select two projects $p_1 \neq p_2 \in P$ randomly.
	Remove all the activities belonging to projects $p_1$ and $p_2$ from the sequence.
	Fill the gaps with all the $p_2$ activities and then all the $p_1$ activities preserving the original order within each of the projects.
	If, in the original solution, $p_1$ was located earlier than $p_2$ then the move swaps the projects (for instance, $(1, 1, 1, 2, 1, 2, 2, 2)$ turns into $(2, 2, 2, 2, 1, 1, 1, 1)$).
	Otherwise it simply separates them more clearly (for instance, $(2, 2, 2, 1, 2, 1, 1, 1)$ turns into $(2, 2, 2, 2, 1, 1, 1, 1)$).

	\item
	\move{Swap neighbour projects}:	swap two projects adjacent in the project ordering.
	Extract the project ordering $P(S)$, randomly select $1 \le i < q$ and set $p_1 = P(S)_i$ and $p_2 = P(S)_{i+1}$.
	Then apply the \move{Swap two projects} move.

	\item
	\move{Compress project}: place all the activities of a project adjacently to a new location $x$ in the sequence while preserving their original ordering.
	Randomly select a project $p \in P$ and remove all the activities $j \in P_p$ from the sequence, squeezing the gaps (the resulting sequence will contain $n - |P_p|$ activities).
	Insert all the activities $j \in P_p$ consecutively at position $\lceil x (n - |P_p|) \rceil$, where the parameter $0 \le x \le 1$ is the new relative location of $P_p$.
	


	\item 
	\move{Shift project}: shift all the activities of a project by some offset.
	Randomly select a project $p \in P$ and calculate $\mathit{pos}_\text{min} = \min_{j \in P_p} \mathit{pos}_j$ and $\mathit{pos}_\text{max} = \max_{j \in P_p} \mathit{pos}_j$.
	Randomly select an offset $-\mathit{pos}_\text{min} < \delta n - \mathit{pos}_\text{max}$ and shift every activity $j \in P_p$ by $\delta$ positions toward the end of the sequence.

	\item 
	\move{Flush projects}: flush the activities of one or several projects to the beginning or ending of the sequence.
	Compute the project ordering $P(S)$ and select $x$ consecutive projects $P'$ from $P(S)$, where $1 \le x < |P|$ is a parameter.
	Flush all the activities in projects $P$ to either the beginning or the ending of the sequence (defined by an additional parameter of the move).
	
\end{itemize}

\myclearpage
\section{Improvement Phase}
\label{sec:imp}

 Most of the time our algorithm spends on improving the initial solutions.
 We use a multi-threaded implementation of a simple memetic algorithm with a powerful local search procedure based on a hyper-heuristic which controls moves discussed in the previous section.

\ignore{
In the improvement phase, moves are controlled by a combination of a meta-heuristic and a hyper-heuristic, all in the context of a multi-threaded population-based approach that uses ideas from memetic algorithms.}

\subsection{Memetic Algorithm}
\label{sec:memetic}

 A genetic algorithm is a population based metaheuristic combining principles of natural evolution and genetics for problem solving \citep{Sastry2014}.  
 A pool of candidate solutions (individuals) for a given problem is evolved to obtain a high quality solution at the end. 
 A fitness function is used to measure the quality of each solution. 
 Mate/parent selection, recombination, mutation and replacement are the main operators of an evolutionary algorithm. 
 However, the usefulness of recombination is still under debate in the research community \citep{Doerr:2008,mitchell93}. 
 A recent study showed that recombination can be useful at a certain stage during the search process, if the mutations do not change the quality of resultant individuals leading to a population containing different individuals with the same fitness \citep{Sudholt:2012}.
 The choice of the recombination operator can influence the best setting for the rate of mutation depending on the problem dealt with.
 Although the study is rigorous, it is still limited considering that some benchmark functions, such as {\sc OneMax} are used for the proofs.
 A memetic algorithm (MA) hybridises a genetic algorithm with local search which is commonly applied after mutation on the new individuals \citep{Moscato:89,Moscato:92}. 
 Many improvements for MAs have been suggested, for example the population sizing \citep{Karapetyan2011} and interleaved mode of operation \citep{Ozcan:MA:2012}. 
 MAs have been successfully applied to many different problems ranging from generalised travelling salesman \citep{Gutin2009gtsp-memetic} to nurse rostering \citep{Ozcan2007}.
 
 The improvement phase of our algorithm is controlled by a simple multi-threaded MA\@ which manages the solution pool and effectively utilises all the cores of the CPU\@.
 Our MA is based on quantitative adaptation at a local level according to the classification in \cite{Ong:2006}.

 Within the MA, we use a powerful local search procedure that takes a few seconds on a single core to converge to a good local minimum.
 (Running the local search for a longer time still improves the solution but the pace of improvements slows down.)
 As the local search procedure has to be applied to every solution of the population in each generation, and the local search is by far the most intensive time consumer in our algorithm, the total running time of the algorithm can be estimated as 
 $$
 \frac{\varname{num-gen} \cdot \varname{pop-size} \cdot \varname{ls-time}}{\varname{cores}} \,,
 $$
 where $\varname{num-gen}$ is the number of generations, $\varname{pop-size}$ is the size of the population, $\varname{ls-time}$ is the time taken by the local search procedure on one core per solution and $\varname{cores}$ is the number of available CPU cores.
 Our algorithm is designed to converge within a few minutes.
 This implies that, to have a sufficient number of generations, the size of the population ought to be of the same order as the number of cores.
 On the other hand, as the memetic algorithm requires a barrier synchronisation at each generation, and our local search procedure cannot utilise any more than one CPU core, the size of the population has to be $\varname{pop-size} = \varname{cores} \cdot i$ to fully utilise all the cores, where $i$ is a positive integer.
 In our implementation, we fixed the running time of the local search procedure to five seconds, and the size of the population to the number of CPU cores available to the algorithm.

 Because of the small population size and limited number of generations, we decided to use a simple version of the MA, see Algorithm~\ref{alg:memetic}.
 We denote the population as $\mathcal{S} = \{ S_1, S_2, \ldots, S_{\varname{cores}} \}$, and the subroutines used in the algorithm are as follows:

\begin{itemize}
	\item 
	$\mathit{Construct}()$ returns a new random solution with the initial partial project sequence obtained during the construction phase (see Section~\ref{sec:constructor}).
	
	\item 
	$\mathit{Accept}(S_i)$ returns $\mathit{true}$ if the solution $S_i$ is considered `promising' and $\mathit{false}$ otherwise.
	The function returns $\mathit{false}$ in two cases: (1) $f(S_i) > 1.05 f(S_{i'})$ for some $i' \in \{ 1, 2, \ldots, |\mathcal{S}| \}$ or (2) the solution was created at least three generations ago and $S_i$ is among the worst three solutions.
	Ranking of solutions is performed according to $f_\text{d}(S_i) + \mathit{idle}$, where $\mathit{idle}$ is the number of consecutive generations that did not improve the solution $S_i$.
	
	\item 
	$\mathit{Select}(S)$ returns a solution from the population chosen with the tournament selection based on two randomly picked individuals.
	
	\item 
	$\mathit{Mutate}(X)$ returns a new solution produced from solution $X$ by applying a mutation operator.  
	The mutation operator to be applied is selected randomly and uniformly among the available options:
	\begin{itemize}
		\item Apply the \move{Reshuffle positions and modes} move with the \move{Local} selection strategy (see Section~\ref{sec:moves-rnr}).
		The cardinality of the selection is $|A'| = 3$.
		Repeat the procedure 20 times, each time randomly selecting the centre of distribution $1 \le \tau \le f_\text{m}(D)$, see (\ref{eq:move-local}).  
		The $\varname{width}$ parameter is taken as $\varname{width} = 0.1 f_\text{m}(D)$.
		
		\item Apply the \move{Swap neighbour projects} operator once (see Section~\ref{sec:moves-project-level}).
		
		\item Apply the \move{Flush projects} operator with the number of selected projects being one and flushing to the end of the sequence (see Section~\ref{sec:moves-project-level}).
		
		\item Apply the \move{Flush projects} operator with the number of selected projects being two and flushing to the beginning of the sequence.
		It was noted that the project ordering is usually less explicit at the beginning of the sequence in good solutions, and for that reason we did not use flushing single projects to the beginning in our mutations.

		\item Same as the last mutation except that the number of selected projects is three.
	\end{itemize}
\end{itemize}

 It was noted in Section~\ref{sec:constructor} that the components $f_\text{d}(D)$ and $f_\text{m}(D)$ of the objective function $f(D)$ (see (\ref{eq:f})) are competing.
 Indeed, minimisation of the total makespan favours solutions with projects running in parallel as such solutions are more likely to achieve higher utilisation of the global resources.  
 At the same time, minimisation of the TPD favours solutions with the activities grouped by projects.  
 Hence, the second objective creates a pressure for the local search that pushes the solutions away from the local minima with regards to the first (main) objective.  
 To avoid this effect, we initially disable the second objective ($\gamma \gets 0$, see Section~\ref{sec:pr}) and re-enable it only after 70\% of the given time is elapsed.


 The parameters of the memetic algorithm (such as the ones used in the $\mathit{Accept}(S_i)$ function, or the number of solutions in the tournament in $\mathit{Select}(S)$) have been chosen using a parameter tuning procedure.
 It should be noted, however, that the algorithm is not very sensitive to any of those parameters, which makes it efficient on a wide range of instances, and this conclusion was supported by our empirical tests, see Section~\ref{sec:expt}.

\begin{algorithm2e}
\caption{Improvement Phase.}
\label{alg:memetic}
$\gamma \gets 0$\;
\For {$i \gets 1, 2, \ldots, |\mathcal{S}|$ } 
{
	$S_i \gets \mathit{Construct}()$\;
}
\While {$\varname{elapsed-time} \le \varname{given-time}$}
{
	\If {$\varname{elapsed-time} \ge 0.7 \varname{given-time}$}
	{
		$\gamma \gets 0.000001$ (enable secondary objective function)\;
	}
	
	\For {$i \gets 1, 2, \ldots, |\mathcal{S}|$ (multi-threaded)}
	{
		$S_i \gets \mathit{LocalSearch}(S_i)$\;
	}
	
	\For {$i \gets 1, 2, \ldots, |\mathcal{S}|$}
	{
		\If {$\mathit{Accept}(S_i) = \mathit{false}$}
		{
			$X \gets \mathit{Select}(\mathcal{S})$\;
			$S_i \gets \mathit{Mutate}(X)$\;
		}
	}
}
\end{algorithm2e}

\subsection{A Dominance based Hyper-heuristic Using an Adaptive Threshold Move Acceptance}

There is a growing interest towards self-configuring, self-tuning, adaptive and automated search methodologies. 
Hyper-heuristics are such high level approaches which explore the space of heuristics (i.e.\ move operators) rather than the solutions in problem solving \citep{burkeSurvey}. 
There are two common types of hyper-heuristics in the scientific literature \citep{Burke:09}:
{\em selection} methodologies that choose/mix heuristics from a set of preset low-level heuristics (which can both improve or worsen the solution) and attempt to control those heuristics during the search process; and {\em generation} methodologies that aim to build new heuristics from a set of preset components.
%
The main constituents of an iterative selection hyper-heuristic are {\em heuristic selection}  and {\em move acceptance} methods.
At each step, an input solution is modified using a selected heuristic from a set of low-level heuristics. 
Then the move acceptance method is used to decide whether to accept or reject the new solution. 
More on different types of hyper-heuristics, their components and application domains can be found in \citep{burkeSurvey,Burke:et,Ross:05,Ozcan:08}.
%

\begin{sloppypar}
In this study, we combine two selection hyper-heuristics under a single point based iterated local search framework by employing them adaptively in a staged manner. The approach extends the heuristic selection and move acceptance methods introduced in \cite{KheiriDom} and \cite{Hysst}, respectively.
%
%
The pseudocode of our adaptive iterated multi-stage hyper-heuristic  is given in Algorithm~\ref{algo:localSearch}.  lines 5--21 and 22--25 
provide the high level design of the first and second stage hyper-heuristics, respectively.  
\end{sloppypar}

\begin{algorithm2e}
\caption{$\varname{LocalSearch}(S_i)$}
\label{algo:localSearch}

\BlankLine
Let $\varname{LLH}_\text{all}=\{\varname{LLH}_1, \varname{LLH}_2, \ldots, \varname{LLH}_\mathcal{M}\}$ represent set of all low level heuristics with each heuristic being associated with a score, initially set to 1\;
Let $S_\text{best}$ represent the best schedule\;
 $S \gets S_i; S_\text{best} \gets S_i; \varname{LLH} \gets \varname{LLH}_\text{all}; \epsilon \gets \epsilon(S_\text{best})$\;
\Repeat {$\mathit{timeLimitExceeded}(\varname{elapsed-time}_1)$}
{

	$h \gets \mathit{SelectLowLevelHeuristic}(LLH$)\;

	$S' \gets h(S)$\;

	\If {$f(S') < f(S)$}
	{

		$S \gets \mathit{S'}$\;

		\If {$f(S') < f(S_\text{best})$}
		{

			$S_\text{best} \gets S'$\;

		}
	}

	\Else
	{
		\If {$f(S') < (1+\epsilon)f(S_\text{best})$}
		{

			$S \gets S'$\;

		}

	}

    \If  { $\mathit{NoImprovement}(\varname{elapsed-time}_2)$}
	{

		$S \gets \mathit{S_\text{best}}$\;

		$\epsilon \gets \epsilon(S_\text{best})$\;

	}

	\If  {$\mathit{NoImprovement}(\varname{elapsed-time}_3)$}
	{

		$\epsilon \gets \epsilon(S_\text{best})$\;

		$(S, \varname{LLH}) \gets \mathit{SecondStage}(S_\text{best},\varname{LLH}_\text{all},$ $\varname{elapsed-time}_1)$\;

	}

}

\Return $S_\text{best}$\;
\end{algorithm2e}

\subsubsection{First stage hyper-heuristic}
\label{subsectionStage1}

The first stage hyper-heuristic maintains an active pool of low-level heuristics $\varname{LLH} \subseteq \varname{LLH}_\text{all}$ and a score $\varname{score}_h$ associated with each heuristic $h \in \varname{LLH}$.
In each iteration, it randomly selects a low-level heuristic from the active pool with probability of picking $h \in \varname{LLH}$ being proportional to $\varname{score}_h$ (line 5).
Then the selected heuristic is applied to the current solution (line 6). Initially, each heuristic has a score of 1, hence the selection probability of a heuristic is equally likely.
The first stage hyper-heuristic always maintains the best solution found so far, denoted as $S_\text{best}$ (lines 9--11) and keeps track of the time since last improvement.

\begin{sloppypar}
The move acceptance component of this hyper-heuristic (lines 7--17) is an adaptive threshold acceptance method controlled by a parameter $\epsilon$ accepting all improving moves (lines 7--12) and sometimes non-improving moves (lines 14--16).
If the quality of a new solution $S'$ is better than $(1 + \epsilon) f(S_\text{best})$  (line 14), even if this is a non-improving move, $S'$ gets accepted becoming the current solution $S$.
Whenever $S_\text{best}$ can no longer be improved for $\varname{elapsed-time}_2$ (in our implementation $\varname{elapsed-time}_2 = 1 \text{ sec}$), the parameter $\epsilon$ gets updated as follows:
\begin{equation}
	\label{eq:epsilon}
	\epsilon(S_\text{best}) = \frac{\lceil \log(f(S_\text{best}))\rceil+\mathit{rand}}{f(S_\text{best})}
\end{equation}
where $1 \le \mathit{rand} \le \lceil \log(x)\rceil$ is selected randomly at uniform.
Note that 0 is a lower bound for $f(S)$ (see Section~\ref{sec:pr}) and, hence, the algorithm will terminate if $f(S_\text{best}) = 0$.
\end{sloppypar}




\subsubsection{Second stage hyper-heuristic}

The second stage hyper-heuristic dynamically starts operating (lines 22--25 of Algorithm \ref{algo:localSearch}) whenever there is no improvement in $f(S_\text{best})$ for $\varname{elapsed-time}_3$ (in our implementation $\varname{elapsed-time}_3 = 3 \text{ sec}$) in line~22. 
The hyper-heuristic in this stage updates the active pool $\varname{LLH}$ of heuristics.
$\varname{LLH} \subseteq \varname{LLH_\text{all}}$ is formed based on the idea of a dominance-based heuristic selection as introduced in~\cite{KheiriDom} reflecting the trade-off between the objective value achieved by each low-level heuristic and number of steps involved. 
The method considers that a low-level heuristic producing a solution with a small improvement in a small number of steps has a similar performance to a low level heuristic generating a large improvement in large number of steps. This hyper-heuristic not only attempts to reduce the set of low-level heuristics but also assigns a score for each low-level heuristic in the reduced set, dynamically. Those scores are used as a basis for the heuristic selection probability of each low level heuristic to be used in the first stage hyper-heuristic. 

In the second stage hyper-heuristic, firstly, $\epsilon$ is updated in the same manner as in the first stage hyper-heuristic and never gets changed during this phase. 
%
Then a greedy strategy is employed using all heuristics in $\varname{LLH}_\text{all}$ for a certain number of steps. Each move in this stage is accepted using the same adaptive threshold acceptance method as described in Section \ref{subsectionStage1}. 
$\varname{LLH}_\text{all}$ is partitioned into three subsets $\varname{LLH}_\text{small}$, $\varname{LLH}_\text{medium}$, $\varname{LLH}_\text{large}$ considering the number of activities processed (e.g., number of swaps) by a given heuristic.  

\noindent
\textbf{Small}: \move{Swap activities}, \move{Shift} and \move{Change mode}.

\noindent
\textbf{Medium}:
\begin{itemize}
	\item 
	\move{Reshuffle modes} and \move{Reshuffle positions and modes} with the \move{Uniform} selection strategy;
	
	\item
	\move{Reshuffle modes} and \move{Reshuffle positions and modes} with the \move{Local} selection strategy;
	
	\item
	\move{Reshuffle positions and modes} with the \move{Global resource driven} selection strategy;

	\item
	\move{Reshuffle positions and modes} with the \move{End biased} selection strategy;

	\item
	\move{Reshuffle positions and modes} with the \move{Project} selection strategy;
	
	\item
	\move{FILS swap activities}, \move{FILS shift} and \move{FILS change mode};
\end{itemize}

\noindent
\textbf{Large}:
\begin{itemize}
	\item
	\move{Flush projects} applied to one project; the direction is picked randomly;

	\item
	\move{Swap two projects};
	
	\item
	\move{Compress project};
	
	\item
	\move{Shift project}.
\end{itemize}

At each step, each low-level heuristic is applied to the same input solution for a fixed number of iterations ($5n/q$ for small, $n/q$ for medium and 1 for large heuristics).\footnote{Similar to the memetic algorithm parameters, this partition of $\mathit{LLH}_{\mathit{all}}$ and the associated numbers of iterations were selected by intuition and parameter tuning.}
If a low-level heuristic produces a new solution identical to the input, that invocation is ignored. 
Otherwise, the objective of the new solution together with the low-level heuristic which produced that solution gets recorded. 
Once all heuristics are applied to the same input and get processed at a given step, the best resultant solution propagates as input to the next greedy step. 

A solution is considered to be a `non-dominated' solution at a given step, if the quality of the solution is better than the best solution obtained in the previous steps.
The active pool $\varname{LLH}$ of low-level heuristics is formed using the heuristics achieving the non-dominated solutions obtained at each step during the greedy phase.  
The score of a low-level heuristic is set to the number of
non-dominated solutions that it produced. Note that a non-dominated solution could be generated by multiple low-level heuristics, in which case, their scores get increased. 

\begin{figure}[htb]
\centering
\includegraphics[width=0.4\textwidth]{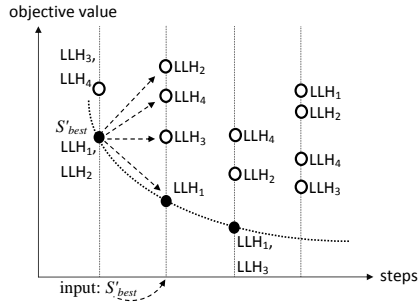}
\caption{An illustration of how the second stage greedy hyper-heuristic operates.}
\label{fig:hh-dom}
\end{figure}

\begin{sloppypar}
Figure~\ref{fig:hh-dom} illustrates a run of the second stage hyper-heuristic with $\varname{LLH}_\text{all}$ = \{$\varname{LLH}_1$, $\varname{LLH}_2$, $\varname{LLH}_3$, $\varname{LLH}_4$\} for four steps. 
The plot shows that the set of non-dominated solutions contains 3 points marked in black.
 The best solution  achieved by $\varname{LLH}_3$ in the last step is ignored, since that solution is dominated by the first three best solutions obtained in the first three steps.  
The first, second and third points in the set of non-dominated solutions  are associated with \{$\varname{LLH}_1, \varname{LLH}_2$\}, \{$\varname{LLH}_1$\} and \{$\varname{LLH}_1,\varname{LLH}_3$\}, respectively. 
Hence, $\varname{LLH}$ becomes \{$\varname{LLH}_1$, $\varname{LLH}_2$, $\varname{LLH}_3$\}. 
The scores of $\varname{LLH}_1$, $\varname{LLH}_2$ and $\varname{LLH}_3$ are assigned to 3, 1 and 1, respectively, yielding a heuristic selection probability of 60\%, 20\% and 20\% for each heuristic. 
More details on the components of the proposed hyper-heuristic can be found in \cite{Kheiri:2015,KheiriThesis}.
\end{sloppypar}

\myclearpage
\section{Experiments}
\label{sec:expt}

 In this section we report the results of our computational experiments as well as discuss the MISTA 2013 Challenge and its outcomes.
 For additional analysis of the performance of our and other submitted algorithms we refer to \citep{Wauters2014}.

\subsection{Test Instances}
\label{sec:instances}

 In our experiments, we used three sets of MRCPSP instances provided by the organisers of the MISTA 2013 Challenge.
 The first set (set A) was provided at the beginning of the competition.
 The second set (set B) was provided after the qualification phase.
 The third set (set X) was hidden from the participants of the competition until the announcement of the results; it was used to evaluate the performance of the submitted algorithms.
 While designing and tuning our algorithm, we had no information on what properties would the hidden instances have, which drove our decisions towards self-adaptive methods and rich range of moves.
 
 \begin{sloppypar}
 All the three sets of instances were produced by combining PSPLIB benchmark instances \citep{Kolisch1997} and can be downloaded from \url{allserv.kahosl.be/mista2013challenge}.
 In Table~\ref{tab:insChar} we provide some properties of all the A, B and X instances.
 \end{sloppypar}

\begin{table*} 
\begin{center}
\scalebox{0.75}{
\begin{tabular}{@{}l *{15}{r} @{}}
\toprule
Instance 
 & $q$ 
 & $n$ 
 & \shortstack{avg\\$d$}
 & \shortstack{avg\\$|\mathscr{M}|$} 
 & \shortstack{avg\\$|\mathit{Pred}|$}
 & \shortstack{avg\\$|\mathscr{R}^{\rho}|$}
 & \shortstack{avg\\$|\mathscr{R}^{\nu}|$}
 & $|\mathscr{G}^{\rho}|$
 & \shortstack{avg\\$R^{\rho}$}
 & \shortstack{avg\\$C^{\nu}$}
 & \shortstack{avg\\$G^{\rho}$}
 & \shortstack{avg\\CPD}
 & $H$ 
 & \shortstack{max\\height} 
 & \shortstack{avg\\SP} \\
\midrule
A-1	&	2	&	20	&	5.53	&	3	&	1.20	&	1	&	2	&	1	&	18.5	&	51.3	&	16.0	&	14.5	&	167	&	4	&	0.33	\\
A-2	&	2	&	40	&	4.63	&	3	&	1.70	&	1	&	2	&	1	&	23.5	&	117.3	&	23.0	&	22.5	&	303	&	8	&	0.34	\\
A-3	&	2	&	60	&	5.51	&	3	&	1.73	&	1	&	2	&	1	&	38.5	&	154.8	&	49.0	&	33.5	&	498	&	9	&	0.26	\\
A-4	&	5	&	50	&	4.37	&	3	&	1.20	&	1	&	2	&	1	&	15.2	&	44.9	&	12.0	&	14.2	&	409	&	5	&	0.40	\\
A-5	&	5	&	100	&	5.43	&	3	&	1.70	&	1	&	2	&	1	&	24.0	&	92.4	&	13.0	&	23.0	&	844	&	8	&	0.32	\\
A-6	&	5	&	150	&	5.13	&	3	&	1.73	&	1	&	2	&	1	&	23.8	&	175.4	&	13.0	&	25.6	&	1166	&	9	&	0.26 \\
A-7	&	10	&	100	&	6.03	&	3	&	1.20	&	0	&	2	&	2	&	0.0	&	48.4	&	11.5	&	16.8	&	787	&	6	&	0.37	\\
A-8	&	10	&	200	&	5.67	&	3	&	1.70	&	0	&	2	&	2	&	0.0	&	110.8	&	22.5	&	24.6	&	1569	&	9	&	0.31\\
A-9	&	10	&	300	&	5.61	&	3	&	1.73	&	1	&	2	&	1	&	27.5	&	168.0	&	27.0	&	29.6	&	2353	&	11	&	0.27\\
A-10	&	10	&	300	&	5.53	&	3	&	1.73	&	1	&	2	&	1	&	25.9	&	158.2	&	15.0	&	30.7	&	2472	&	10	&	0.26\\[1ex]
B-1	&	10	&	100	&	5.33	&	3	&	1.20	&	1	&	2	&	1	&	17.1	&	44.8	&	11.0	&	12.9	&	821	&	5	&	0.34\\
B-2	&	10	&	200	&	5.67	&	3	&	1.70	&	0	&	2	&	2	&	0.0	&	94.0	&	21.0	&	23.9	&	1628	&	7	&	0.28\\
B-3	&	10	&	300	&	5.52	&	3	&	1.73	&	1	&	2	&	1	&	28.5	&	144.4	&	28.0	&	29.5	&	2391	&	10	&	0.26\\
B-4	&	15	&	150	&	5.03	&	3	&	1.20	&	1	&	2	&	1	&	17.5	&	52.3	&	10.0	&	15.8	&	1216	&	6	&	0.36\\
B-5	&	15	&	300	&	6.02	&	3	&	1.70	&	1	&	2	&	1	&	20.7	&	99.6	&	17.0	&	22.5	&	2363	&	8	&	0.31\\
B-6	&	15	&	450	&	4.62	&	3	&	1.73	&	1	&	2	&	1	&	25.0	&	141.8	&	34.0	&	31.1	&	3582	&	11	&	0.26\\
B-7	&	20	&	200	&	4.87	&	3	&	1.20	&	1	&	2	&	1	&	14.7	&	49.6	&	10.0	&	15.4	&	1596	&	5	&	0.37\\
B-8	&	20	&	400	&	5.48	&	3	&	1.70	&	0	&	2	&	2	&	0.0	&	104.7	&	10.0	&	23.7	&	3163	&	8	&	0.30\\
B-9	&	20	&	600	&	5.31	&	3	&	1.73	&	1	&	2	&	1	&	26.6	&	154.8	&	10.0	&	30.1	&	4825	&	12	&	0.26\\
B-10	&	20	&	420	&	5.28	&	3	&	1.66	&	0	&	2	&	2	&	0.0	&	115.9	&	18.0	&	24.5	&	3340	&	12	&	0.31\\[1ex]
X-1	&	10	&	100	&	5.53	&	3	&	1.20	&	0	&	2	&	2	&	0.0	&	47.6	&	12.5	&	14.9	&	783	&	5	&	0.36\\
X-2	&	10	&	200	&	5.53	&	3	&	1.70	&	1	&	2	&	1	&	24.0	&	105.6	&	14.0	&	23.0	&	1588	&	8	&	0.32\\
X-3	&	10	&	300	&	4.98	&	3	&	1.73	&	1	&	2	&	1	&	27.9	&	167.0	&	33.0	&	29.9	&	2404	&	11	&	0.27\\
X-4	&	15	&	150	&	5.70	&	3	&	1.20	&	0	&	2	&	2	&	0.0	&	54.5	&	13.5	&	14.9	&	1204	&	5	&	0.36\\
X-5	&	15	&	300	&	5.52	&	3	&	1.70	&	1	&	2	&	1	&	19.9	&	100.1	&	12.0	&	23.6	&	2360	&	8	&	0.32\\
X-6	&	15	&	450	&	5.49	&	3	&	1.73	&	1	&	2	&	1	&	24.6	&	163.7	&	20.0	&	29.9	&	3597	&	10	&	0.26\\
X-7	&	20	&	200	&	5.03	&	3	&	1.20	&	1	&	2	&	1	&	13.9	&	53.9	&	10.0	&	15.0	&	1542	&	5	&	0.33\\
X-8	&	20	&	400	&	5.53	&	3	&	1.70	&	1	&	2	&	1	&	22.2	&	104.5	&	15.0	&	24.5	&	3217	&	8	&	0.32\\
X-9	&	20	&	600	&	5.54	&	3	&	1.73	&	1	&	2	&	1	&	23.9	&	146.3	&	11.0	&	28.9	&	4699	&	10	&	0.26\\
X-10	&	20	&	410	&	5.30	&	3	&	1.65	&	1	&	2	&	1	&	20.0	&	101.2	&	10.0	&	24.1	&	3221	&	9	&	0.30\\
\bottomrule
\end{tabular}
}
\caption{Characteristics of the instances: 
$q$ is the number of projects,
$n$ is the number of activities, 
avg $d$ is the average duration of activities in all possible modes, 
avg $|\mathscr{M}|$ is the average number of modes for each activity, 
avg $|\mathit{Pred}|$ is the average number of the predecessor activities for each activity, 
avg $|\mathscr{R}^{\rho}|$ is the average number of the local renewable resources per project, 
avg $|\mathscr{R}^{\nu}|$ is the average number of the local non-renewable resources per project, 
$|\mathscr{G}^{\rho}|$ is the number of global renewable resources, 
avg $R^{\rho}$ is the average renewable resource capacities, 
avg $C^{\nu}$ is the average non-renewable resource capacities, 
avg $G^{\rho}$ is the average global renewable resource capacities, 
avg $\text{CPD}$ is the average CPD per project,
$H$ is the upper bound on the time horizon. 
The last two columns give an indication of the properties of the network (precedence graph); 
max height is the length of the longest path, e.g.\ see \url{https://users.soe.ucsc.edu/~manfred/pubs/J2.pdf} 
and avg SP is a measure of the `series vs. parallel' nature \citep{Vanhoucke2013:overview}.
}
\label{tab:insChar}
\end{center}
\end{table*}

\myclearpage
\subsection{Experimental Results}
\label{sec:results}


 During the competition, the submitted algorithms were tested by running each of the algorithms on each of the B and X instances 10 times.
 The results for X (hidden) instances were used to rank the teams.
 The detailed discussion of the ranking scheme and the analysis of the competition results are reported in \citep{Wauters2014}.

 To further analyse the performance of our algorithm, we ran it for 5 minutes on each of the B and X instances 2500 times.
 We used a machine with similar configuration\footnote{Intel i7 3.2~GHz CPU, 16 GB of RAM, Microsoft Windows 7 x64 operating system} to that employed by the competition organisers.
 
 The aggregated results of our experiment are reported in Table~\ref{tab:experiments}.
 The average and best solutions are obtained from our 2500 runs, while the competition (``Comp'' column) result is the best solution found during the final phase of the competition by our or some other algorithm.

 

 It is worth mentioning here that during the final phase of the competition where the submitted algorithms were tested on B and X instances, our approach outperformed, on average, all the other algorithms in 18 out of 20 cases, and produced the best solution for 17 instances.
 \cite{Wauters2014} note the consistency of the performance of our algorithm, and also provide ranks with several different ranking methodologies, showing that our approach would win the competition even if the rules of the competition were different.

\begin{table}[bth!]
\begin{center}
\begin{tabular}{@{}l @{\qquad} r r r r r r r @{}}
\toprule
& \multicolumn{3}{c}{TPD} & \quad & \multicolumn{3}{c}{TMS} 	\\
\cmidrule(r){2-4}
\cmidrule(){6-8} 	
Instance & Avg  & Comp  & Best  && Avg  & Comp  & Best    \\ \midrule
A-1 	& 1		& 1		& 1 	&& 23	&	23	& 23	\\
A-2 	& 2		& 2		& 2 	&& 41	&	41	& 41	\\
A-3 	& 0		& 0		& 0 	&& 50	&	50	& 50	\\
A-4 	& 65	& 65	& 65 	&& 42	&	42	& 42	\\
A-5 	& 155	& 153	& 150 	&& 105	&	105	&103 	\\
A-6 	& 141	& 147	& 133 	&& 808	&	96	&99 	\\
A-7 	& 605	& 596	& 590 	&& 201	&	196	&190 	\\
A-8 	& 292	& 302	& 272 	&& 153	&	155	&148 	\\
A-9 	& 208	& 223	& 197 	&& 128	&	119	&122 	\\
A-10 	& 880	& 969	& 836 	&& 313	&	314	&303 	\\[1ex]

B-1 	& 352	& 349	& 345 	&& 128	&	127	&124 	\\
B-2 	& 452	& 434	& 431 	&& 167	&	160	&158 	\\
B-3 	& 554	& 545	& 526 	&& 210	&	210	&200 	\\
B-4 	& 1299	& 1274	& 1252 	&& 283	&	289	&275 	\\
B-5 	& 832	& 820	& 807 	&& 255	&	254	&245 	\\
B-6 	& 950	& 912	& 905 	&& 232	&	227	&225 	\\
B-7 	& 802	& 792	& 782 	&& 232	&	228	&225 	\\
B-8 	& 3323	& 3176	& 3048 	&& 545	&	533	&523 	\\
B-9 	& 4247	& 4192	& 4062 	&& 754	&	746	&738 	\\
B-10 	& 3290	& 3249	& 3140 	&& 455	&	456	&436 	\\[1ex]

X-1 	& 405	& 392	& 386 	&& 143	&	142	&137 	\\
X-2 	& 356	& 349	& 345 	&& 164	&	163	&158 	\\
X-3 	& 329	& 324	& 310 	&& 193	&	192	&187 	\\
X-4 	& 960	& 955	& 907 	&& 209	&	213	&201 	\\
X-5 	& 1785	& 1768	& 1727 	&& 373	&	374	&362 	\\
X-6 	& 730	& 719	& 690 	&& 238	&	232	&226 	\\
X-7 	& 866	& 861	& 831 	&& 233	&	237	&220 	\\
X-8 	& 1256	& 1233	& 1201 	&& 288	&	283	&279 	\\
X-9 	& 3272	& 3268	& 3155 	&& 648	&	643	&632 	\\
X-10 	& 1613	& 1600	& 1573 	&& 383	&	381	&373 	\\
\bottomrule
\end{tabular}
\caption{Summary of experimental results. (Note that the results for the A instances are for the algorithm of the first stage of the competition, and are included only for completeness.)}
\label{tab:experiments}
\end{center}
\end{table}

\begin{figure}[htb]
    \begin{center}
	\includegraphics{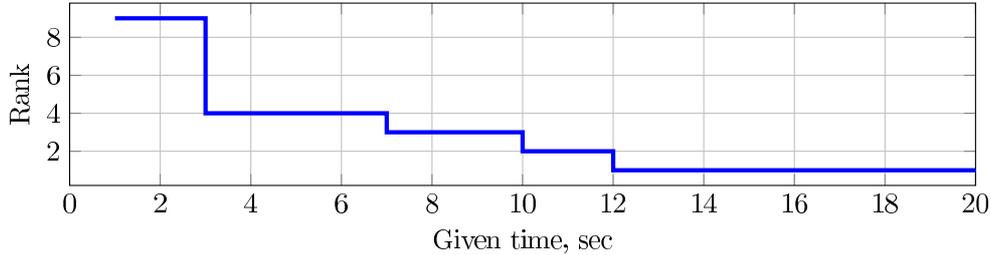}
    \end{center}
\caption{The rank achieved by our algorithm with various time limits in the assumption that the other algorithms are given 300 seconds.  Given just 12 seconds, our algorithm would already achieve the rank 1.}
\label{fig:dynamicrank}
\end{figure}

 
 We conducted an additional experiment to investigate the effect of the runtime on the ranking results, and so an indication of whether or not the code improvements as discussed in Section~\ref{sec:code-opt}, might have been the predominant reason, as opposed to the algorithm itself, of achieving the first rank during the competition.
 In this experiment, we ran our algorithm (in the style which was used during the competition) with reduced time limits.
 After each run, we determined the rank of our algorithm among all other competitors (the data required for this ranking was received from competition organisers). 
 Figure~\ref{fig:dynamicrank} shows the result of the experiment for various time limits ranging from 1 to 25 seconds.
 For the first two or three seconds the algorithm runs MCTS thus not generating high quality solutions (recall that the goal of MCTS is to find a high-quality partial ordering of projects).
 However, as soon as the memetic algorithm starts, the solution quality rapidly improves.
 In fact, our algorithm manages to achieve the first rank after just 12 seconds.
 Given that the time limit during the competition was 300 seconds, this means that the algorithm would have still ranked highest even if it had run 25 times slower. 
 In contrast the improvements of Section~\ref{sec:code-opt} did not give that factor of a speed up.
 Assuming that the other submissions were reasonably implemented then we conclude that efficient implementation was not the only factor, but that the algorithm design also had a major role in the success of the algorithm.

 To illustrate the performance of our approach in time, two boxplots are provided in Figures~\ref{fig:performanceB1} and~\ref{fig:performanceX10} for instances B-1 and X-10, respectively.  
 These instances have been chosen to demonstrate the performance of our algorithm on relatively small and large instances. 
 However, our experiments show that the algorithm behaves in a similar manner with respect to the other instances.
 The central dividing line of each box in each of the figures presents the median objective value obtained by our algorithm. 
 The edges of each box refer to $25^{th}$ and $75^{th}$ percentiles while the whiskers (demonstrated by a $+$ marker) are the extreme objective values which are not considered as outliers. 
 Also, the curve which passes through the plot demonstrates the average performance of the algorithm.
 Our approach continuously makes improvement indicating that the algorithm does not prematurely converge.

\ignore{
\begin{figure*}
\begin{center}
\subfigure{\includegraphics[width=0.7\textwidth]{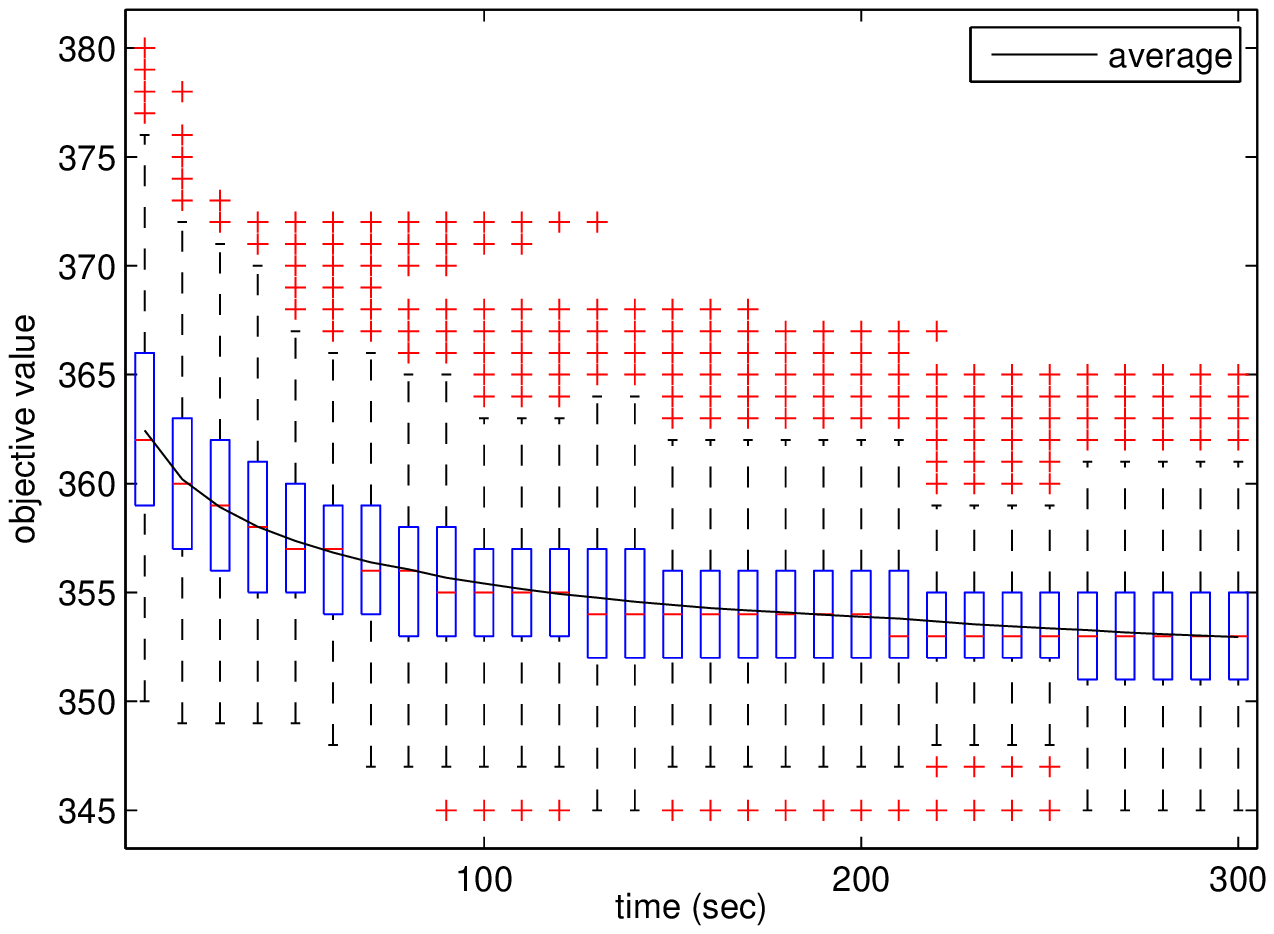}} 
\end{center}
\caption{Performance of the proposed approach on instance B-1.}
\label{fig:performanceB1}
\end{figure*}

\begin{figure*}
\begin{center}
\subfigure{\includegraphics[width=0.7\textwidth]{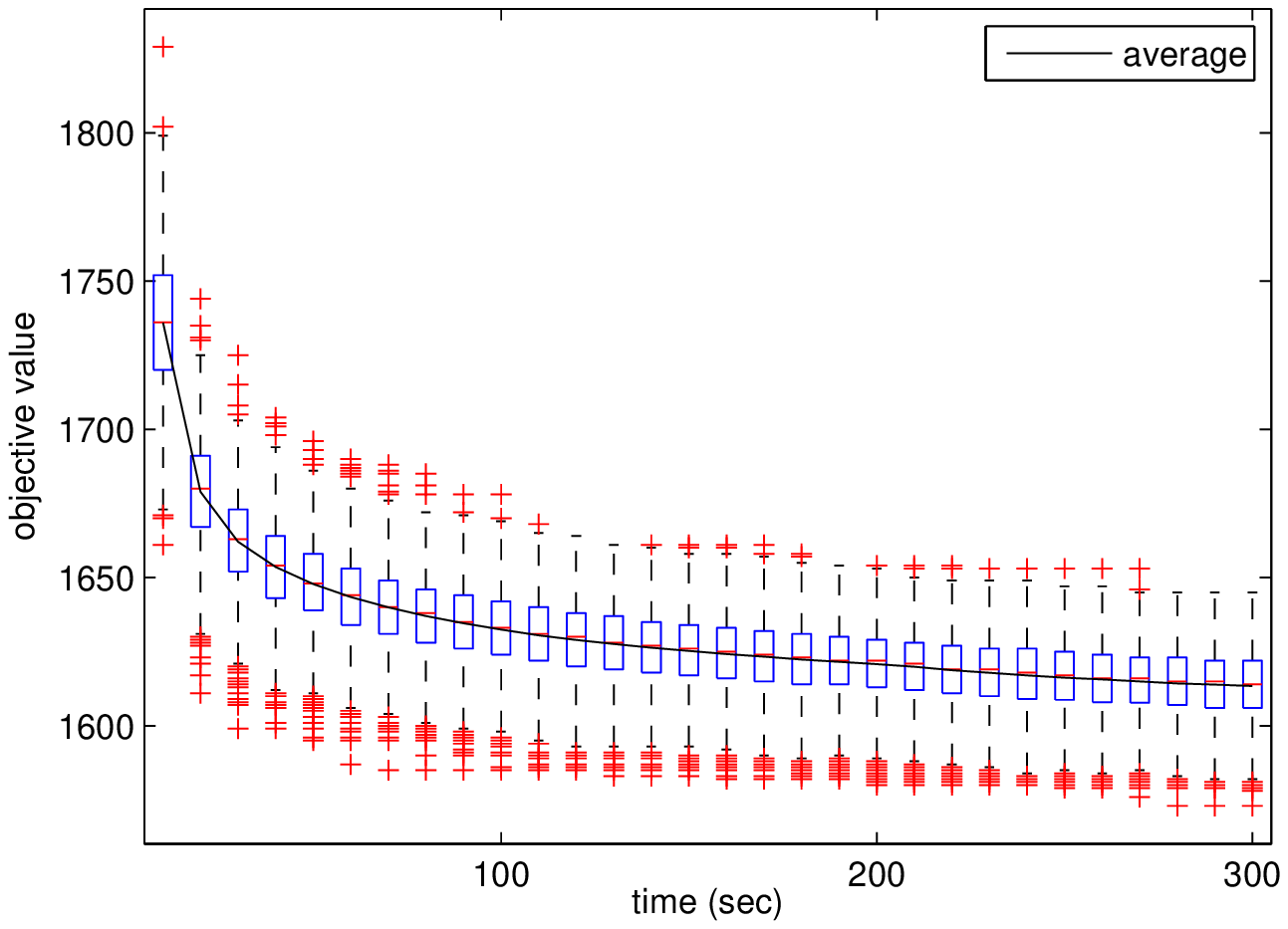}} 
\end{center}
\caption{Performance of the proposed approach on instance X-10.}
\label{fig:performanceX10}
\end{figure*}
}

\begin{figure*}
\begin{center}
\begin{subfigure}[t]{0.49\textwidth}
	\includegraphics[width=\textwidth]{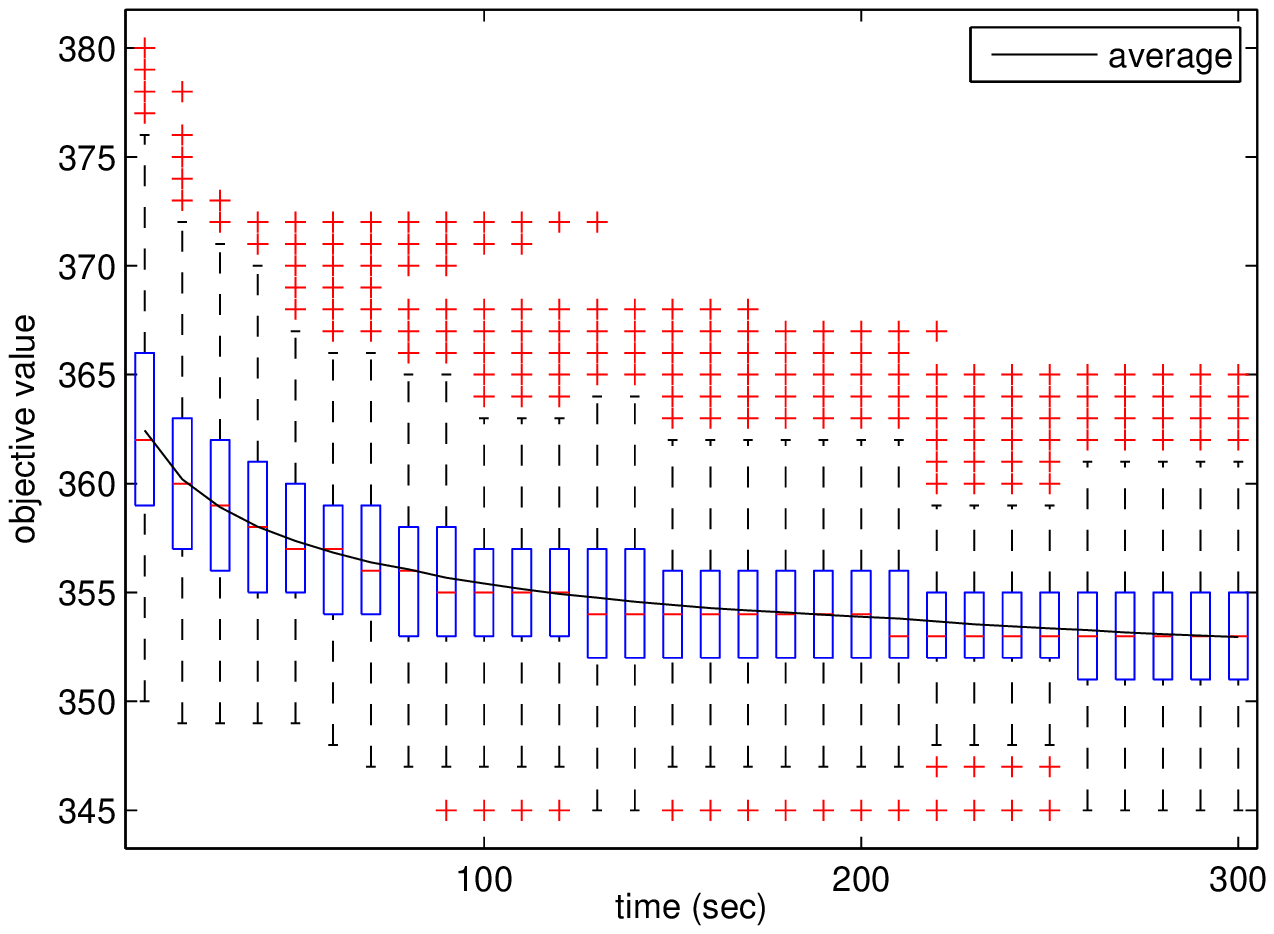}
	\caption{B-1}
	\label{fig:performanceB1}
\end{subfigure}
\begin{subfigure}[t]{0.49\textwidth}
	\includegraphics[width=\textwidth]{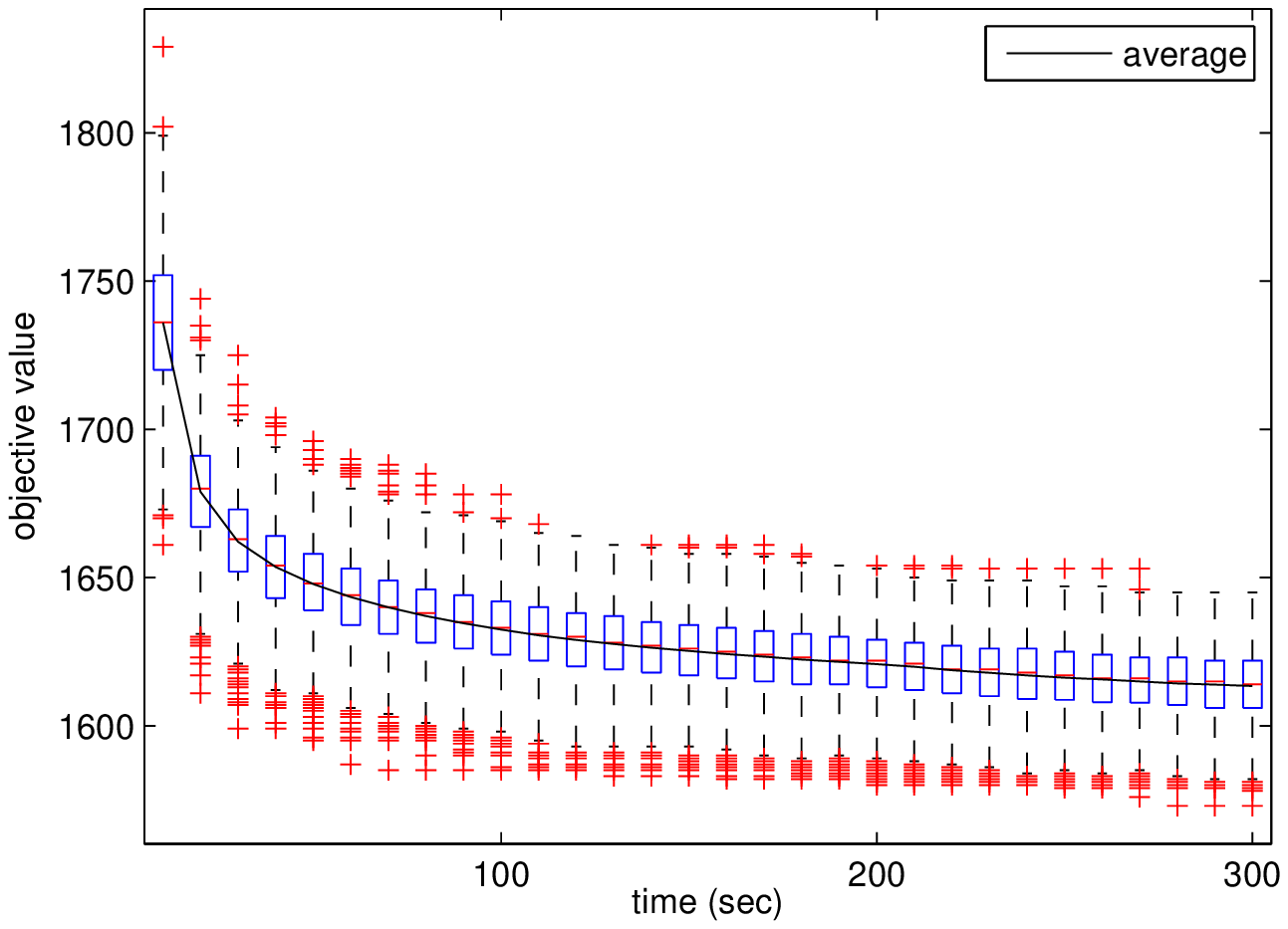}
	\caption{X-10}
	\label{fig:performanceX10}
\end{subfigure}
\end{center}
\caption{Performance dynamics of the proposed approach.  The horizontal axis corresponds to the time spent by the algorithm on optimisation and the vertical axis shows the solution quality.  The results are averaged over 2500 runs and represented in the box plot form.}
\label{fig:performanceB1X10}
\end{figure*}

\begin{sloppypar}
 It would obviously be interesting to link together the performance of the algorithm(s) with the features of the instances; for example, see \cite{Peteghem2011,Messelis2014} and others. 
 There are many potential features that could be used to characterise the instances, and these have often been used as part of the process of generating instances, for example \citep{Kolisch1995,Demeulemeester2003}.
 For example, one can characterise the precedence graph in various ways \citep{Reyck1996,Kamburowski2000,Emmert-Streib2012} and associated measures such as `complexity' and `Series-Parallel' indices could also be included.
%
%
 This would be far outside the scope of this paper, and warrants a future study. 
%
%
 However, we can make some initial observations by analysing the instance properties in Table~\ref{tab:insChar} and the results in Table~\ref{tab:experiments}, and in more detail in Table 33 of the final competition summary \citep{Wauters2009}.
 In particular, we focus on the property of the number of local renewable resources in the column ``avg $|\mathscr{R}^{\rho}|$'', and that there are none for instances B-2, B-8, B-10, X-1 and X-4.
 These seem to correlate with the performance in that the algorithm here only fails to rank first on instances B-2 and X-1, both of which have no local renewable resources. 
 We remark that having no local renewables is a rather special case, and so if these were removed the relative performance would have further improved.
 However, it seems reasonably likely that the number of local renewable resources would be good instance feature for selecting algorithms, or for future studies comparing different algorithms in more details.
\end{sloppypar}

\ignore{

Note we do not play the common game of implicitly setting many parameters to have the same value and then claiming that the algorithm has few parameters.
Not that would necessarily be a bad idea, but because at this stage we do not understand the parameter space well enough to be able to do this in a sensible fashion.

We hope that future research will elucidate this, and presumably will also incorporate machine learning to the tuning

TODO: include a table of

for each A and B instance

- average of the best 5 minute algorithm

- best from the quals phase (for the A instances)

- best solution we ever saw in any run with any algorithm or parameters  -- just useful for reference

NOTE: we should try to do some long runs to get some decent best-known solutions?

TODO Asta: add several plots, update the tables, ensure that the averages in the tables are calculated based on the first objective only, provide some discussion of the results.

TODO Asta: as far as I can see, there is no description of the machines we were testing our algorithm on.  Please mention it.

}

\myclearpage

\subsection{Results on Single Project (PSPLIB) Instances}
\label{sec:psplib}

The main aim of this work is to develop methods for the MISTA benchmark problems having multiple projects.
However, it is also of interest to study the methods on other scheduling benchmarks; in particular, the PSPLIB library\footnote{\url{http://www.om-db.wi.tum.de/psplib/}} of instances \cite{Kolisch1997} has been widely used and studied in the academic scheduling community.
Unfortunately, despite it being long known that the majority of real world problems are multi-project \citep{Lova2000}, it seems that much of the work in project scheduling has been on single project instances, and PSPLIB does not currently contain any multi-project instances.
Despite this limitation of PSPLIB, it has functioned well as a standard set of benchmarks, and so it is likely to interest how well the methods of this paper perform on these instances.
The most relevant 
set of instances in PSPLIB is provided by the multi-mode single project instances\footnote{\url{http://www.om-db.wi.tum.de/psplib/getdata.cgi?mode=mm}}.
The study of those instances is also limited considering the number of tasks which
goes up to 30 tasks, whereas the MISTA competition included instances with up to 600 tasks. 
(Other PSPLIB instances are either much simpler as they are just single mode, or else introduce extra constraints of objectives for the timing, and so are outside the scope of this current paper.)
Specifically, we studied the J30 instances from PSPLIB; they all have precisely 3 modes for each of the 30 tasks, with 2 renewable and 2 non-renewable resources, and we use the 552 of them that are satisfiable (a feasible solution is known). 
%

For the special case of single project, minimising the TPD objective becomes the same as minimising the makespan, TMS\@.
Hence, the PSPLIB multi-mode instances can be regarded as a special single-project case of the MISTA formulation and so our methods could be directly applied to these instances to minimise the makespan.
However, doing so would be rather inefficient as many portions of our methods are designed to handle multiple-projects and the associated TPD objective.
Hence, we partially specialised our code in order to handle these simpler PSPLIB instances. 
In particular, we inactivated the Monte-Carlo Tree Search as it is only there in order to handle the structures arising from the TPD objective, replacing it with a simple random generation heuristic. 
We also inactivated the neighbourhoods that are specific to working on a specific project.
Since making comparisons to previous work also needed a very limited time period and just a single thread, we also disabled the memetic component as it was designed to exploit multi-cores and longer runs; to partially compensate for the resulting expected loss in diversity in the search, we also added a simple restart of the search every 10k generated schedules.


These instances have been studied in various works.
%
%
\cite{PeteghemVanhoucke2011:mRCPSP-GA}
used a Genetic algorithm and also Forward-Backward Iterations (FBI) \citep{LiWillis1992:forward-backward} along with a simple method to repair infeasible mode choices. 
\cite{CoelhoVanhoucke2011:multi-mode-RCPSP-SAT}
again used FBI along with a more expensive SAT-based method to ensure that only feasible mode decisions were made 
(recently, \cite{PeteghemVanhoucke2014:mRCPSP} reported that this was the best method on these instances).
In these works, the standard fashion is to impose a limit on the number of schedules generated.
The reported performance measure is then the gap, averaged over all instances, from either a known lower bound or else from the current ``Best Known Solutions (BKS)''. 
Often, the number of schedules permitted is rather small, such a 5--10k, which would take our code less than a second, and so would not really match the intent of our methods. 
Hence, in Table~\ref{tab:J30} we compare with previous results at 50k and 500k from Table~4 of \cite{CoelhoVanhoucke2011:multi-mode-RCPSP-SAT}. 
We also note that with 500k schedules, and under 2s, our heuristic found the best-known solutions in more than 85\% of instances, and in 99\% it was within 1 time unit of the BKS\@.
We also performed much longer runs to exploit the full power of all the algorithm components: up to 500M generated schedules within the concurrent memetic, though still only taking up to about 400 seconds, due to the concurrency, and the more efficient implementations (see Section~\ref{sec:generator}).
We were able to find 3 new best solutions; these have been added to PSPLIB, and so giving the only improvements to these multi-mode J30 instances since 2011.

\begin{table}[tbh!]
\begin{center}
\begin{tabular}{@{}l @{\qquad} rrrr @{}}
\toprule
Schedules generated:
	& \multicolumn{2}{@{}r}{50k}				
	& \multicolumn{2}{@{}r@{}}{500k} \\
\cmidrule(r){2-3}
\cmidrule(l){4-5}
	& Gap, \%
	& Time, sec
	& Gap, \%
	& Time, sec\\ 
\midrule
This work					 		
	& 13.68
	& 0.12
	& 12.84
	& 1.14\\
Coelho and Vanhoucke (2011)
	& 12.77
	& 25.1\phantom{1}
	& 12.41
	& 210.1\phantom{1}\\
Van Peteghem and Vanhoucke (2010) 	
	&  13.31         	
	&  2.46
	& 13.09
	& 18.03\\
\bottomrule
\end{tabular}
\caption{Summary of experimental results on PSPLIB multi-mode J30 instances for 50k and 500k generated schedules. Giving the average gap from the lower bound (the lower the better) and the average running time (the lower the better). Note that the runtimes are only indicative as the older results will be on machines that might well be 2x slower.
}
\label{tab:J30}
\end{center}
\end{table}

Hence, we conclude that the methods here perform competitively on these multi-mode instances. 
We believe this gives evidence of the power and flexibility of the overall approach. 
This is particularly noteworthy, as our methods were originally designed for the more general multi-project case with the extra delay based objective, which has quite different properties, as already discussed in Section~\ref{sec:constructor}; the difference between minimising TPD and TMS can lead to a significant difference in the nature of the solutions, and reasonable algorithms.
For example, both of the compared approaches \citep{PeteghemVanhoucke2011:mRCPSP-GA,CoelhoVanhoucke2011:multi-mode-RCPSP-SAT} used the successful forward-backward iterations, however our methods for did not include it, because it is not naturally applicable to the primary objective being TPD\@.
Including it in our suite of improvement operators, might well improve performance for the special case of improving makespan.

Finally, we remark that recent work on these multi-mode single project instances by \cite{Messelis2014} has considered the problem of selecting algorithms. 
This is perhaps the closest in spirit to our goals, in that the aim is to take multiple options for algorithms and then to use intelligent or machine learning techniques in order to make the selection. 
In contrast to a more traditional approach in which the exploration of the combinations of many components is done manually.

\myclearpage

\section{Conclusions}
\label{sec:conc}

\begin{sloppypar}
 This paper described the components of a multi-component hybrid approach to the MRCMPSP\@. 
 Broadly, our algorithm is built on the serial generation scheme, in which sequences of activities are used to construct final schedules for their quality evaluation.
 We use a two-phase construct-and-improve method to search the space of activity sequences, with a variety of novel contributions, including algorithm mechanisms specifically designed to handle the multi-project structure and associated objectives functions.
\end{sloppypar}
 
 Firstly, the primary objective is a sum of completion times of the individual projects, and we give evidence and arguments that this tends to lead to an imbalance in the completion times; some projects finish much earlier than others. 
 Hence, the nature of the problem itself is such that an approximate partial ordering of projects occurs in good solutions, and this is quite different from the structure expected with makespan as the primary objective. 
 We expect such approximate partial ordering may well be common in real-world multi-project problems as well.
 Accordingly, in the construction phase, a novel Monte-Carlo Tree Search method is given that creates and selects initial solutions with an approximately-similar structure.

 Secondly, in the improvement phase, multiple novel neighbourhoods are included that are designed to work at the `project-level' and so enable any needed changing of the approximate partial ordering of the projects.
 These are complemented with a wide range of neighbourhood moves also designed to work at lower levels within projects.
 The improvement phase itself is organised and controlled using a novel hybrid of a memetic algorithm and a hyper-heuristic, and in a fashion that furthermore makes effective use of a multi-core machine when it is available.


 Perhaps a distinguishing characteristic of the methods here is that there are far more neighbourhoods than in other submissions to the competition, or indeed the literature in general. 
 From Table~6 in \citep{Wauters2014}, other submissions have around 1--4 neighbourhoods but we have 13--17 (though many neighbourhoods have different variants, and are used in different stages, and so there is not a single meaningful count).
 At first this might seem to be a step towards extra complexity, and that might make the algorithm more difficult to use in real-world practice.
 However, in many problem domains there is a drive towards a larger number of neighbourhoods, for example, in realistic educational timetabling where 10--15 neighbourhoods is not uncommon (for example \citep[and others]{Hysst}).
 Having many neighbourhoods is a natural reaction to the increasing complexity of the structure of the problem.
 They are designed to work on each level of the roughly hierarchical structure of many projects with each project containing many activities. 
 We only eliminated neighbourhoods in a stage when they clearly never helped; instead preferring to keep neighbourhoods, with the intention that the diversity would help in robustness and effectiveness (though admittedly at the cost of `neatness').
 Also, the objective itself is a type of weighted completion time and so is sensitive to the internal structure of solutions. 
 Hence, it seems reasonable that many neighbourhoods are needed; and this seems to be be borne out by the much better performance.
 Furthermore, in the case of scheduling using the serial generation there is an advantage and opportunity in that the neighbourhoods are mostly `lightweight' to implement.
 They work by modifying the activity sequence, and when the majority of the runtime is spent on the evaluation of a sequence, then designing and efficiently implementing many different neighbourhoods is much less onerous then many other problem domains.

 Another significant difference to other submissions to the challenge, and as described in~\citep{Wauters2014}, it seems that many others in the competition used forward-backward iteration (FBI) methods.
 We did not use it due to the extra complexity, and also uncertainty of how to best use it for the TPD (weighted completion time) rather than the standard makespan.
 Also, the results in \citep{PeteghemVanhoucke2014:mRCPSP} said it did not ``distinguish between a good or bad performing metaheuristic''.
 However, it would be interesting future work to also include FBI moves; possibly with appropriate modification for the TPD objective.

The effectiveness of the resulting hybrid algorithm was demonstrated when it convincingly won the MISTA 2013 challenge, outperforming, on average, the other algorithms on 18 of the 20 instances \cite{Wauters2014}.
The many-component and many-neighbourhood structure was a deliberate decision on our part, and was aimed at giving a more robust solver.  
In particular, part of the design goal was that the methods would be robust and so work well on the hidden instances, and it did indeed do particular well on those; ranking first on 9 out of the 10 hidden instances.
 We also observed that the performance was weakest (though still high ranking) on those instances with no local renewable resources. 
 Such instances are rather special, and arguably less realistic, however, it does suggest that there is good potential for future work to investigate the links between features of the instances and the the algorithm performance.
 Nevertheless, Section~\ref{sec:psplib} gives additional evidence of the robustness of our algorithm as even a much-reduced version performs well on the simple single-project but multi-mode instances from PSPLIB.

 The broader contribution of this work is to give evidence for the general approach and methodology of investigating algorithms that consist of many components (such as many neighbourhoods, but also other aspects) and that aims to exploit the potential benefits of simple, but carefully controlled, interactions between them.
 It is possible that the algorithm could be simplified (e.g.\ by reducing the number of neighbourhood moves); however, we suspect that doing so is not worth the risk of it only working well on the seen set of instances, and would lose its robustness and hence performance on the unseen instances.
 Hence, we believe that the work gives evidence that individual components should not be prematurely discarded.
Instead, ultimately, the decision as to which combination works best over a particular suite should be automated; that is, as a form of algorithm assembly, for example in the style of \citep{BezLopStu2014:ppsn}. 
Of course, such assembly can also use a form of algorithm selection for the appropriate scheduling algorithm, based on the features of instances (e.g.\ \cite{Peteghem2011,Messelis2014}).
 
Such `automated assembly' work would be outside the scope of this particular paper; however, by providing a broad range of options and components that have the proven potential to give a successful and robust solver, this work hence gives a good foundation for such studies in the MRCMSP, and other variants of project scheduling, and arguably intelligently-coordinated metaheuristic methods in general.

\bigskip

\noindent
\textbf{Acknowledgements} This work was supported in part by EPSRC Grants EP/F033214/1 and EP/H000968/1.

\ignore{
As an examples
Given existing neighbourhoods, N1 N2 N3    and of which (on some suite) N1 is the most effecttive single nbd
it is quite possible some new nbd N' may beat that best one.  However, could still be outperformed by carefully controlled simple hybrids.
N' might beat N1 but could lose to N1+N2
Ideally, evaluations of potential new neighbourhoods, should take account of this, and show that they provide value not available by combining existing methods (or they do so more effectively).  The design problem is that of selecting sets of operrators that work well together, and this may well be different than comparisons of methods using only sinle (or a few methods).

However, of course, the number of hybrids is combinatorial in the number of operators, and so the number of potential algorithms becomes unmanageable to do directly.  Hence, comparisons and evaluations of methods will need to be able to evaluate against effective hybrid systems. 
This works, firstly provides evidence that such hybrids can be highly effective, also it offers foundational steps forhyper-heuristics to be one framework that can mange such hybrids in an intelligent and dynamic fashion. 

Possibly, most radically, it suggests that emphasis of studies should not necessarily be pre-selection of (small sets of) good operators, but also the ways in which larger sets can be effectively combined.
}

\AJPii{

WRONG QUESTION

On benchmarks Y what is the minimal set of operators, determined "by hand", that minimises avg gap?

RIGHT QUESTION

What is a broad set of operators, and assocaited coordination such that on benchmarks Y it uses a reasonable set, and that gives equivalent results?  With the point being that on switching to new benchmarks it will still work as is more robust.

In the extreme we might say

Our deepest interest lies in how information from the runtime should be used in order to achieve better control and coordination of the components, as opposed to only being interested in the details of the final answer.

but do not take such an extreme.
But do argue about the other extreme in that only the final answer is of interest.

}

\ignore{
\AJP{The points below need to be merged with some of those above to avoid redundancy}

We emphasise that one aim was creating a robust solver that it would have a better chance of dealing with the unseen instances.  
During the design process we did some effort to reduce the set of neighbourhoods, however, it was more to remove potential neighbourhoods that never seemed to be useful. 
As a result, there is no claim (not even implicit) of it being minimal in terms of what is needed for the specific instances studied.  
Of course, one could take the specific set of instances and then try to determine smaller combinations that  perform well on the specific instances. 
This would be informative, and we intend to do this in future work, however, there is a danger in such methodology of resulting in a set of components that are too limited, and are potentially over-tuned to a, limited and known-in-advance, set of instances. 
We note again that on the hidden instances our relative performance to others actually slightly improved.
Instead, the intention is that there are good reasons that we have a robust set of internal components are likely to work well together; in our case with the MCTS and project moves giving good search at the project level, and others working best at the activity level.  
Individual components should not be prematurely discarded;  ultimately, the decision as to which combination works best over a particular suite should be automated; that is, as a form of algorithm assembly, for example in the style of \cite{BezLopStu2014:ppsn}. 
Of course, such assembly can also use a form of algorithm selection for the appropriate scheduling algorithm, based on the features of instances e.g. \cite[and others]{Peteghem2011,Messelis2014}. 
Such work would be outside the scope of this particular paper; however, by providing a broad range of options and components that have the proven potential to give a successful and robust solver, this work hence gives a good foundation for such studies in the MRCMSP, and other variants of project scheduling.
}

\myclearpage

\bibliographystyle{plain}  

\bibliography{bib}   

\end{document}